\documentclass[12pt,psf,epsf]{article}
\usepackage[centertags]{amsmath}
\usepackage{amssymb}
\usepackage{mathtools}
\usepackage{graphicx,dsfont}
\usepackage{epsfig}
\usepackage{ulem}
\usepackage[english]{babel}
\usepackage{array}
\usepackage{amsthm}
\usepackage{latexsym}
\usepackage[mathcal]{euscript}
\pdfoutput=1
\usepackage{epsfig}
\usepackage{jheppubm}
\usepackage{mathdots}
\usepackage{MnSymbol}
\usepackage{multirow}
\usepackage{comment}
\usepackage[export]{adjustbox}

\newcommand{\be}{\begin{equation}}
\newcommand{\ee}{\end{equation}}
\newcommand{\bea}{\begin{eqnarray}}
\newcommand{\eea}{\end{eqnarray}}

\newcommand{\dd}{\partial}
\newcommand{\sgn}{\text{sgn}}

\newcommand{\mO}{\mathcal{O}}

\newcommand{\mH}{\mathcal{H}}

\renewcommand{\i}{{\mathfrak{i}}}
\newcommand{\bZ}{{\bf Z}}
\newcommand{\bI}{{\bf I}}

\newcommand{\bS}{\mathbf{S}}

\DeclareMathAlphabet\mathbfcal{OMS}{cmsy}{b}{n}
\newcommand{\bcZ}{\mathbfcal{Z}}
\newcommand{\bch}{\mathbf{\mathfrak{h}}}
\newcommand{\cZ}{\mathcal{Z}}

\usepackage{graphicx}

\newcommand{\IPL}[1]{I_{\vcenter{\hbox{\includegraphics[width= 2 mm]{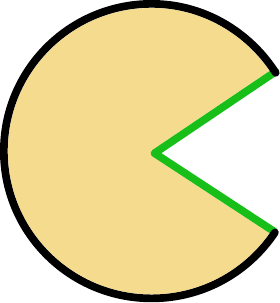}}}#1}} 
\newcommand{\IPiR}[1]{I_{\vcenter{\hbox{\includegraphics[width= 1.6 mm]{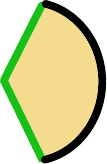}}}#1}} 
\newcommand{\ICL}[1]{I_{\vcenter{\hbox{\includegraphics[width= 1.8mm]{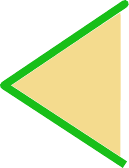}}}#1}}
\newcommand{\ICLB}[1]{I_{\vcenter{\hbox{\includegraphics[width= 1.8mm]{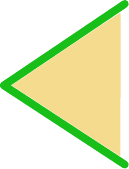}}}#1}}
\newcommand{\ICR}[1]{I_{\vcenter{\hbox{\includegraphics[width= 1.8mm]{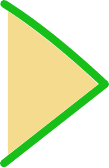}}}#1}}

\newcommand{\IPR}[1]{I_{\vcenter{\hbox{\includegraphics[width= 2mm]{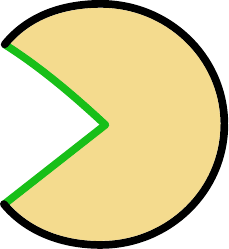}}}#1}}

\newcommand{\PL}{\vcenter{\hbox{\includegraphics[width= 5 mm]{figures/packmanL.pdf}}}}
\newcommand{\PiR}{\vcenter{\hbox{\includegraphics[width= 3.5 mm]{figures/pizzaR.pdf}}}}
\newcommand{\CL}{\vcenter{\hbox{\includegraphics[width= 4mm]{figures/coneL.pdf}}}}
\newcommand{\CLB}{\vcenter{\hbox{\includegraphics[width= 4mm]{figures/coneLblue.pdf}}}}
\newcommand{\SR}{\vcenter{\hbox{\includegraphics[width= 0.3mm]{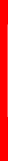}}}}
\newcommand{\SP}{\vcenter{\hbox{\includegraphics[width= 0.3mm]{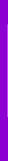}}}} 
\newcommand{\CR}{\vcenter{\hbox{\includegraphics[width= 4mm]{figures/coneR.pdf}}}}

\newcommand{\PR}{\vcenter{\hbox{\includegraphics[width= 5mm]{figures/packmanr.pdf}}}}

\definecolor{dgreen}{rgb}{0.,0.6,0.}
\definecolor{orange}{rgb}{0.91, 0.41, 0.17}
\definecolor{purple}{rgb}{0.6, 0.2, 0.8}

\title{Counting microstates of out-of-equilibrium black hole fluctuations}

\author[1,2,3]{Vijay Balasubramanian,}
\author[2]{Ben Craps,}
\author[2]{Juan Hernandez,}
\author[1,2]{Mikhail Khramtsov,}
\author[2]{Maria Knysh}

\affiliation[1]{David Rittenhouse Laboratory, University of Pennsylvania, 209 S. 33rd Street, Philadelphia, PA 19104, USA}
\affiliation[2]{Theoretische Natuurkunde, Vrije Universiteit Brussel (VUB) and The International Solvay Institutes, Pleinlaan 2, B-1050 Brussels, Belgium}
\affiliation[3]{Rudolf Peierls Centre for Theoretical Physics, University of Oxford, Beecroft Building, Parks Road Oxford OX1 3PU, UK}

\emailAdd{vijay@physics.upenn.edu}
\emailAdd{ben.craps@vub.be}
\emailAdd{juan.hernandez@vub.be}
\emailAdd{mikhail.khramtsov311@gmail.com}
\emailAdd{maria.knysh@vub.be}

\abstract{
We construct and count the microstates of out-of-equilibrium eternal AdS black holes in which a shell carrying an order one fraction of the black hole mass is emitted from the past horizon and re-absorbed at the future horizon.  Our microstates have semiclassical interpretations in terms of matter propagating behind the horizon. We show that they span a Hilbert space with a dimension equal to the exponential of the horizon area of the intermediate black hole.  This is consistent with the idea that, in a non-equilibrium setting, entropy is the logarithm of the number of microscopic configurations consistent with the dynamical macroscopic state. In our case, therefore, the entropy should measure the number of microstates consistent with a large and atypical macroscopic black hole fluctuation due to which part of the early time state becomes fully known to an external observer.
}

\begin{document}

\maketitle
\section{Introduction}

A central idea of quantum statistical mechanics is that the entropy of a macroscopic system is equal to the logarithm of the dimension of the Hilbert space spanned by microstates consistent with the macroscopic specification. For example, the entropy of a gas in equilibrium is derived by counting the ways in which the underlying molecules can be organized given the total number of molecules, the volume of the system, and either the temperature $T$ or total energy $E$, depending on whether we wish to work canonically or microcanonically.  The vast majority of the microscopic configurations contributing to the entropy will be in \textit{typical} states in which most coarse-grained measurements, such as particle density, will give roughly the same results independently of when, where, and in how large a region the measurements are done.  However, a much smaller number of configurations will be \textit{atypical}, in that they will show large deviations, at least transiently, from the ensemble average of measurements. An example of such an atypical microstate of a gas is one in which an order one fraction of the particles transiently accumulates in a corner of the total volume, before spreading back into the bulk.  We can define an entropy associated to transient, {\it macroscopic} fluctuations of this kind by counting how many microstates are consistent with it.  Atypical configurations are always expected at late times, because of large but rare statistical fluctuations that inevitably occur in many-body statistical systems. Such fluctuations exhibit non-trivial time evolution, so understanding their statistical properties is useful for understanding out-of-equilibrium dynamics in statistical systems.

The same issues arise in the thermodynamic description of black holes. The entropy of a black hole is universally given by the Bekenstein-Hawking formula $S_{BH} = A/4G_N\hbar$, where $A$ is the area of the horizon.  An explicit basis that spans the underlying Hilbert space of dimension $e^{S_{BH}}$ was constructed in \cite{shells1,shells2,Climent24,Balasubramanian:2024yxk} for any theory of gravity that has General Relativity as its low-energy classical limit.\footnote{For related work on semiclassical black hole microstates, see also~\cite{Kourkoulou17,Penington19,Chandra:2022fwi,Boruch:2023trc,Boruch:2024kvv,Geng:2024jmm,Li2024,Banerjee2024}.}  Just like gases, black holes can also exhibit macroscopic fluctuations.  The Poincar\'{e} recurrences discussed in~\cite{Bocchieri:1957cnr,Percival:1961,Schulman:1978,Dyson:2002pf} are an example.  As another example, consider an eternal black hole in AdS space that emits from its past horizon a spherical shell of matter or radiation which carries an order one fraction of the total energy of the black hole.   Because of the AdS gravitational potential, the shell will turn back towards the black hole and be reabsorbed.  This situation is parallel to the transient, macroscopic fluctuation of a gas discussed above. In this paper, we will compute the dimension of the Hilbert space spanned by microstates of such atypical, out-of-equilibrium, black hole systems.

Four sections follow.  In~\autoref{sec:setup} we review the construction of semiclassically well-defined microstates for black holes~\cite{shells1,shells2,Climent24,Balasubramanian:2024yxk}.  In~\autoref{sec:review} we discuss how to count the dimension of the Hilbert space of microstates consistent with a given macroscopic geometry outside the horizon.  In~\autoref{sec:mainresult} we apply these methods to out-of-equilibrium eternal black holes which emit and then re-absorb a shell of matter or radiation. For concreteness, the calculations are carried out in AdS$_3$ in this section.  We conclude with a discussion in~\autoref{sec:discussion}, which includes comments on the generalization to higher dimensions. \autoref{app: D=3} contains additional details on the Lorentzian continuation of the microstates. \autoref{sec: orientation} clarifies an important subtlety related to the Euclidean geometries that dominate the gravitational path integral.

 \section{Semiclassical black hole microstates}\label{sec:setup}

We begin by introducing the formalism developed in~\cite{Sasieta:2022ksu,shells1}, with the concrete example of semiclassical microstates with a single shell used in previous work~\cite{shells1,Balasubramanian:2024yxk} for clarity. This will provide the ingredients necessary to reproduce the results of~\cite{shells1} in~\autoref{sec:review} and to build the microstates consistent with atypical black hole fluctuations discussed in the introduction, which we do in~\autoref{sec:mainresult}. We model atypical black hole fluctuations as thin shells of matter or radiation that are emitted and reabsorbed by the black hole, so the tools in this section directly apply to the construction of the corresponding microstates.

\subsection{Definition}
\label{sec: definition}
We consider states defined on a tensor product of two copies of a d-dimensional CFT. Each copy lives on a spacetime of topology $\mathbb{S}^{d-1}\times \mathbb{R}$ and has Hamiltonian $H_L =H_R=H$. The total Hilbert space is $\mathcal{H}=\mathcal{H}^{CFT}_L \otimes \mathcal{H}^{CFT}_R$. The energy basis of this doubled Hilbert space is
\begin{equation}
    (H_L \otimes \mathds{1}) |m \rangle_L \otimes |n \rangle_R=E_m|m \rangle_L \otimes |n \rangle_R\,,\qquad (\mathds{1} \otimes H_R)|m \rangle_L \otimes |n \rangle_R=E_n|m \rangle_L \otimes |n \rangle_R\,.
\end{equation}

\begin{figure}[t]
    \centering
    \includegraphics[scale=0.7]{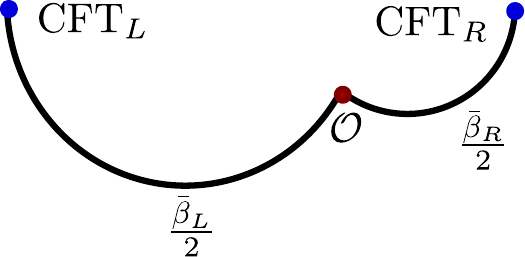} \quad 
    \includegraphics[scale=0.7]{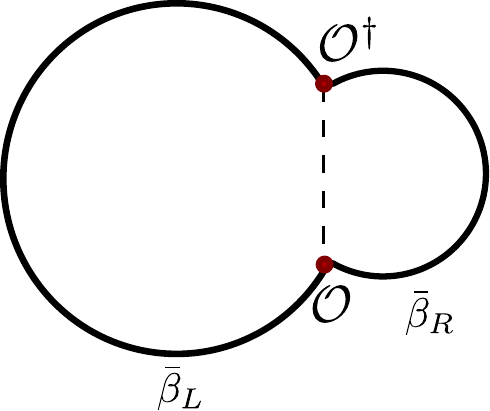} \\ 
    (a) \hspace{5cm} (b)
    \caption{(a) Path integral contour that prepares the state given in eq.~\eqref{state with one shell} The lengths of the circular segments are given by the Euclidean preparation times $\Bar{\beta}_L/2$ and $\Bar{\beta}_R/2$. (b) Euclidean time contour which computes the norm squared of the state given in eq.~\eqref{state with one shell}. This quantity can be computed using the gravitational path integral evaluated for geometries with asymptotic boundary length ${\Bar{\beta}_L + \Bar{\beta}_R}$ and with matter excitations sourced by the operators $\mO$ and $\mO^\dagger$. In both figures, there is a transverse sphere $\mathbb{S}^{d-1}$ at every point.} 
    \label{fig:contouroneshell}
\end{figure}

We consider a family of states constructed via a Euclidean path integral as shown in~\autoref{fig:contouroneshell}(a). To do so, we insert operators $\mO$ that are uniformly supported on a spatial slice of a single copy of the  CFT, and then evolve by the Hamiltonian to the right and left over Euclidean times $\Bar{\beta}_R$ and $\Bar{\beta}_L$. This procedure yields  states of the form
\be
| \Psi \rangle = \frac{1}{\sqrt{Z_\Psi}} \sum_{n, m} e^{-\frac12 \Bar{\beta}_L E_m - \frac12 \Bar{\beta}_R E_n} \mO_{mn} |m \rangle_L \otimes |n \rangle_R\,, \label{state with one shell}
\ee
where $\mO_{mn} = \langle m | \mO | n \rangle$ and $Z_\Psi=\text{Tr}\left[ \mO^\dagger e^{- \Bar{\beta}_L H }\mO e^{-\Bar{\beta}_R H}\right]$ ensures that the state is normalized. The trace in $Z_\Psi$ is taken in a single copy of the CFT. These states are higher dimensional generalizations of the partially entangled
thermal states (PETS) defined in JT gravity~\cite{Goel:2018ubv}.

Using the AdS/CFT correspondence, we can interpret the CFT states constructed in this way as dual to semiclassical spatial wormhole geometries connecting two asymptotically AdS$_{d+1}$ regions. These wormholes contain specific configurations of matter induced by the operator insertions~\cite{Sasieta:2022ksu}. Below, we will focus on operators $\mO$ that are a product of $n$ scalars $\mO_\Delta$ with conformal dimension $\Delta$ that are uniformly supported on the $(d-1)$-dimensional spherical time slice of the CFT. We take ${\cal O}_\Delta$ to be the HKLL reconstruction of bulk operators which insert particles of mass $m_\Delta$ a distance $\epsilon$ away from the asymptotic boundary~\cite{Hamilton:2006az}.\footnote{The mass $m_\Delta$ determined from the standard relation $m_\Delta^2 = \Delta (\Delta - d)$, and throughout this paper, we set the AdS length $L_{\rm AdS}=1$.} Thus, in the dual gravitational theory, the composite operators $\mO$ create spherically symmetric thin shells of matter of mass $m=n m_\Delta$, and we pick $n=O\left(G_N^{-1}\right)$. This scaling ensures that the shell generated by $\mO$ is sufficiently heavy to induce classical backreaction on the geometry at leading order in Newton’s constant, $G_N$, which can be computed using the thin shell formalism~\cite{Israel:1966rt}. The normalization constant $Z_\Psi$ can then be computed using the gravitational path integral with the appropriate boundary conditions, as explained in~\autoref{fig:contouroneshell}.

\subsection{Spherically symmetric thin shells in AdS }
\label{sec: shells}

We first review the dynamics of thin shells~\cite{Israel:1966rt} on a Euclidean geometry that is symmetric under Euclidean time reversal. In the thin shell approximation, the shell worldvolume divides Euclidean spacetime into two regions, which we will call $X^+$ and $X^-$, each satisfying the vacuum Einstein equations.\footnote{In this section, the labels $\pm$ indicate generic regions of the geometry on either side of the shell of matter. In the context of the geometries appearing in later sections, the labels $\pm$ will be replaced by $L,\i$, and $R$ depending on which region they correspond to.} Assuming that the asymptotic temperature at each boundary exceeds the Hawking-Page transition temperature~\cite{Hawking:1982dh}, each region is part of a Euclidean black hole solution rather than of a thermal AdS spacetime.\footnote{This assumption was not specified in previous work on semiclassical microstates~\cite{shells1,shells2,Climent24,Balasubramanian:2024yxk}. However, those authors avoided the subtlety by using a microcanonical projection after which black hole geometries dominated.} In each region, the  black hole metric is
\be\label{eq: blackholemetric}
ds_\pm^2 =g^\pm_{\mu\nu} dx^\mu dx^\nu= f_\pm(r) d\tau_\pm^2 + \frac{dr^2}{f_\pm(r)} + r^2 d\Omega_{d-1}^2\,,
\ee
where
\bea 
\label{eq: fpm}
f_\pm(r) = \begin{cases} 
    r^2 + 1 - \frac{16\pi G M_\pm}{(d-1) V_\Omega r^{d-2}}, & \text{for } d > 2,\\
    r^2 - 8\pi G M_\pm, & \text{for } d = 2.
\end{cases}
\eea
Here, $V_\Omega$ is the volume of a unit sphere $\text{Vol}\left(\textbf{S}^{d-1}\right)=\frac{2\pi^{d/2}}{\Gamma\left(d/2\right)}$. The black hole regions have horizons at $r_\pm$ with corresponding inverse temperatures $\beta_\pm = \frac{4\pi}{f'_\pm(r_\pm)}$. 

We have to glue these black hole geometries along the shell by using the Israel junction conditions~\cite{Israel:1966rt}. Let us define the intrinsic coordinates $\xi^a$, with $a = 0, \dots, d$ on the worldvolume embedded into the two black hole regions. The shell worldvolume inherits induced metrics from the two black hole regions, defined as $h^\pm_{ab} = \frac{\dd x^\mu}{\dd \xi^a} \frac{\dd x^\nu}{\dd \xi^b} g_{\mu\nu}^\pm$ and extrinsic curvatures defined as $K_{ab}^\pm = \frac{\dd x^\mu}{\dd \xi^a} \frac{\dd x^\nu}{\dd \xi^b} \nabla_\mu n^\pm_\nu$, where $n^\mu_\pm$ is an outwards pointing unit normal vector. We define
\bea
\Delta h_{ab} &=& h_{ab}^+ - h_{ab}^-\,,\\
\Delta K_{ab} &=& K_{ab}^+ - K_{ab}^-\,,\\
\Delta K &=& h^{+,ab} K_{ab}^+ - h^{-,ab} K_{ab}^-\,.
\eea
The Israel junction conditions that determine the trajectory of the shell worldvolumes are
\bea
&& \Delta h_{ab} = 0\,, \label{Israel-1}\\
&& \Delta K_{ab} - h_{ab} \Delta K = -8 \pi G S_{ab}\,, \label{Israel-2}
\eea
where $S_{ab}=-\sigma u_a u_b$ is the energy momentum tensor of the shell, $\sigma$ is the surface density and $u^a$ is the proper fluid velocity tangent to the worldvolume of the shell. The mass of the shell is conserved along the worldvolume and is given by 
\be\label{eq: sigma}
m = \sigma V_\Omega R^{d-1}\,.
\ee
For spherically symmetric shells, the shell worldvolume can be parameterized by
\be
r = R(T)\,, \quad \tau = \tau_\pm(T)\,,
\ee
where $T$ is the synchronous proper time\footnote{Meaning that the time component of the induced metric is one, $h_{TT}=1$. } along the shell worldvolume.
The full set of equations of motion of the shell which follows from eqs.~\eqref{Israel-1}-\eqref{Israel-2}, reads
\bea
&&f_\pm^2 \dot{\tau}_\pm^2 = -\dot{R}^2 + f_\pm\,, \label{shell-EOM-1}\\
&& -f_+ \dot{\tau}_+ + f_- \dot{\tau}_- = \frac{8 \pi G m}{(d-1) V_\Omega R^{d-2}}\,.\label{shell-EOM-2}
\eea
By substituting~\eqref{shell-EOM-1} into~\eqref{shell-EOM-2} and squaring the resulting equation, we get the shell equation of motion in the form of particle energy conservation law
\be
\dot{R}^2 + V_{\text{eff}} (R) = 0\,, \label{shell-EOM}
\ee
where 
\be
\label{eq: Veff}
V_{\text{eff}}(R) = -f_+(R) + \Gamma(R)^2\,, \quad \Gamma(R) \equiv \frac{M_+ - M_-}{m} -\frac{4\pi G m}{(d-1) V_\Omega R^{d-2}}\,.
\ee
The effective potential $V_{\text{eff}}(R)$ has a turning point at $R = R_*$ defined by the condition $V_{\rm eff}(R_*) = 0$. As an explicit example, consider AdS$_3$. The effective potential in this case is given by
\be
V_{\text{eff}}(R) = -(r^2 - R_*^2)\,, \quad (\text{for} \ d=2)\,,\label{Veff3d}
\ee
where the turning point is located at
\be
R_* = \sqrt{r_+^2 + \left(\frac{M_+ - M_-}{m} -2 G m\right)^2} \,, \quad (\text{for } d=2)\,.\label{R*3d}
\ee

In terms of the shell equation of motion~\eqref{shell-EOM}, all dependence on the shell’s properties, including its mass $m$, is encapsulated in the function $\Gamma(R)$ which has the properties
\be
\lim_{m \to 0} \Gamma(R) = +\infty\,, \qquad \lim_{m \to +\infty} \Gamma(R) = -\infty\,, \quad R \geq r_+\,.
\ee
There is a critical value $m_0$  of the shell mass such that
\bea
\Gamma(R_*) &=& 0\,, \qquad R_* = r_+\, .
\eea
This value is given by
\bea 
m_0^2 = \frac{M_+ - M_-}{4\pi G} (d-1) V_\Omega r_+^{d-2}\,. \label{eq: m_0}
\eea
In the special case $d=2$, $\Gamma(R)$ is independent of $R$ and $m_0^2 = \frac{M_+ - M_-}{2G}$. For a shell mass $m < m_0$, $\Gamma > 0$ along the entire shell worldline, whereas if $m > m_0$, then $\Gamma < 0$ for any $R$. For $d > 2$, however, $\Gamma(R)$ may not have a definite sign throughout the trajectory.

A consequence of squaring \eqref{shell-EOM-2} to get~\eqref{shell-EOM} is that the latter is agnostic to the sign of $\Gamma(R)$ along the trajectory. This sign, however, is physically meaningful and is set by the relation between the shell mass and the difference $M_+ - M_-$. Without squaring, the original equations~\eqref{shell-EOM-1}-\eqref{shell-EOM-2} imply
\be
\sgn(\dot{\tau}_+)\sqrt{f_+ - \dot{R}^2} = \Gamma(R)\,. \label{eq: sign-dtau}
\ee
This expression indicates that the sign of $\Gamma(R)$, determined by the shell mass $m$, directly corresponds to the sign of $\dot{\tau}_+(R)$. Consequently, once the direction of proper time along the shell worldvolume and the mass of the shell are specified, the geometry of the resulting spacetime solution is uniquely determined, as we will explain below. This distinction is essential for analyzing the shell trajectory and matching conditions between regions.

With the equation of motion now derived in terms of the effective potential~\eqref{shell-EOM}, we can explicitly determine the trajectory of the shell in the Euclidean disk by eliminating the proper time $T$ from the Israel junction conditions~\eqref{shell-EOM-1} and~\eqref{shell-EOM-2}. This leads to the relation
\be
d\tau_\pm = \frac{dR}{f_\pm} \sqrt{\frac{f_\pm + V_{\text{eff}}}{-V_{\text{eff}}}}\,. \label{eq: dtau}
\ee
Integrating this equation gives the Euclidean time elapsed along the trajectory of the shell.

An essential aspect of this setup is that in Euclidean signature, the shell can either excise or preserve the center of the Euclidean disk in the ``$+$" and ``$-$" regions after the gluing procedure. After analytic continuation to the Lorentzian signature, the center of the ``$+$" (``$-$") region being excised by gluing translates to the shell being outside the $r_+$ ($r_-$) event horizon. Conversely, if the center of the ``$+$" (``$-$") region remains after gluing, the shell is inside the $r_+$ ($r_-$) event horizon. Depending on which one of these possibilities is realized, the relation~\eqref{eq: sign-dtau} also fixes the sign of $\Gamma$ near $R_*$ through the sign of $\dot{\tau}$ in the ``$+$" and ``$-$" regions. Below, we explore these configurations in more detail. Without loss of generality, we assume that the Euclidean $r_-$-horizon remains unexcised by the shell, meaning the shell’s worldvolume passes to the right of the center of the ``$-$"-disk. 

\paragraph{Shell inside the $r_+$-horizon.}

\begin{figure}[t]
    \centering
    \includegraphics[scale=0.90]{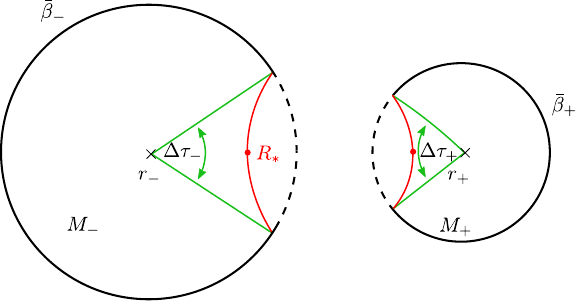} \hspace{1cm}
    \includegraphics[scale=0.60]{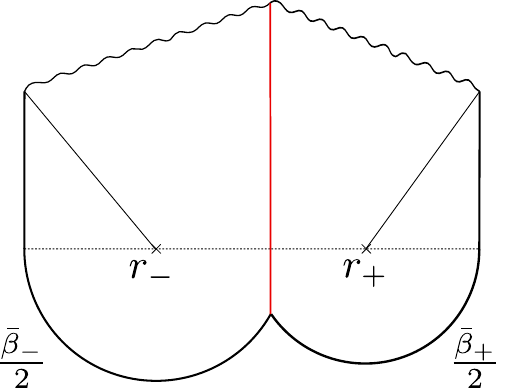} \\
    (a) \hspace{7cm} (b)
    \caption{(a) The trajectory (red line) of the massive thin shell in Euclidean AdS$_3$, governed by eqs.~\eqref{eq: shell-inside-minus}-\eqref{eq: shell-inside-plus}. The crosses denote horizons $r_-$ and $r_+$. The ``$+$'' and ``$-$ regions are glued along the shell trajectory. The dashed curves outline the excised part of the spacetime. Notice that the horizons are not excised, so that after Lorentzian continuation the shell ends up behind both horizons $r_-$ and $r_+$ as shown in (b). }
    \label{fig:shellinside}
\end{figure}

In terms of the shell worldvolume, the choice of the shell being inside or outside the horizons amounts to choosing the boundary condition at the turning point $R_*$. Specifically, for a shell that is inside the $r_+$ horizon, we choose $\tau_+(R_*) = \frac{\beta_+}{2}$ and $\tau_-(R_*) =0$. With these boundary conditions, one can integrate~\eqref{eq: dtau} to find the elapsed time for an internal shell, meaning that it is behind both $r_-$ and $r_+$ horizons,
\bea
\tau_-(R) &=& \int_{R_*}^{R}  \frac{d\Tilde{R}}{f_-} \sqrt{\frac{f_- + V_{\text{eff}}}{-V_{\text{eff}}}}\,, 
\label{eq: shell-inside-minus}\\
\tau_+(R) &=& \frac{\beta_+}{2} - \int_{R_*}^{R}  \frac{d\Tilde{R}}{f_+} \sqrt{\frac{f_+ + V_{\text{eff}}}{-V_{\text{eff}}}}\,. \label{eq: shell-inside-plus}
\eea

As an example, \autoref{fig:shellinside} shows the spacetime with the shell inside the horizons when $d=2$. In this case, the trajectories~\eqref{eq: shell-inside-minus} and \eqref{eq: shell-inside-plus} can be solved,
\bea
\tau_-(R) &=& \frac{\beta_-}{2\pi}\arctan\left(\frac{r_- \sqrt{R^2-R_*^2}}{R\sqrt{R_*^2-r_-^2}}\right)\,, \quad (\text{for } d=2)\,,
\label{eq: shell-inside-minus-d2}\\
\tau_+(R) &=& \frac{\beta_+}{2} -  \frac{\beta_+}{2\pi}\arctan\left(\frac{r_+ \sqrt{R^2-R^2_*}}{R\sqrt{R_*^2-r_+^2}}\right)\,,\quad (\text{for } d=2)\,. \label{eq: shell-inside-plus-d2}
\eea
The boundaries of the ``$+$" and ``$-$" geometries are curves of length $\Bar{\beta}_\pm$, which are related to $\beta_\pm$ as
\be
\beta_+ = \Delta \tau_+ + \Bar{\beta}_+\,,\qquad \beta_- = \Delta \tau_- + \Bar{\beta}_-\,,
\ee
where $\Delta \tau_\pm$ is the total elapsed Euclidean time along the shell worldvolume,
\be
\begin{aligned}
\Delta \tau_\pm & = 2 \int_{R_*}^\infty \frac{d\Tilde R}{f_\pm}\sqrt{\frac{f_\pm + V_{\text{eff}} }{-V_{\text{eff}}}} \\ & = \frac{\beta_\pm}{\pi} \arcsin\left(\frac{r_\pm}{R_*}\right)\,, \quad ({\rm for} \ d=2)\,, \label{DeltaTau}
\end{aligned}
\ee
where the first equality is true in general dimensions, and we have provided the results for $d=2$ in the second line as a specific example. We have taken the shell to start at and return to $R=\infty$. 

Assuming the proper time $T$ is flowing upwards in~\autoref{fig:shellinside}, we have 
\be
\sgn( \dot{\tau_-}) > 0\,, \qquad \sgn( \dot{\tau_+}) < 0\,,
\ee
where recall that for $d=2$ the sign of $\dot{\tau}$ is constant along the entire shell worldvolume. Eq.~\eqref{eq: sign-dtau} then implies that 
\be
\Gamma(R) |_{d=2}= \frac{M_+ - M_-}{m} -2 G m < 0\,.
\ee
This implies that the solution with the shell \textbf{inside} both $r_-$ and $r_+$ horizons is only consistent when the mass of the shell is \textbf{higher} than the critical value, $m > m_0$, where $m_0$ is defined in eq.~\eqref{eq: m_0}.

\paragraph{Shell outside the $r_+$-horizon.}

\begin{figure}[t]
    \centering
    \includegraphics[scale=0.80]{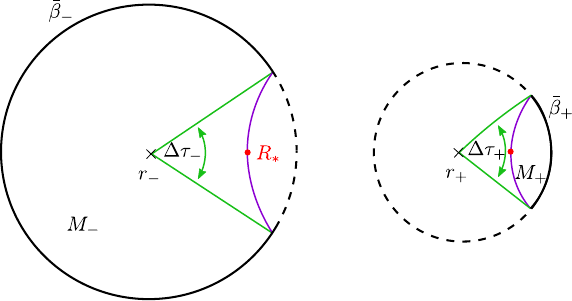} \hspace{0.4cm}
    \includegraphics[scale=0.65]{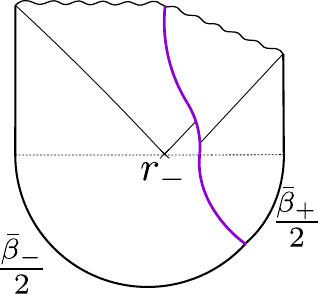} 
    \hspace{0.4cm}
    \includegraphics[scale=0.65]{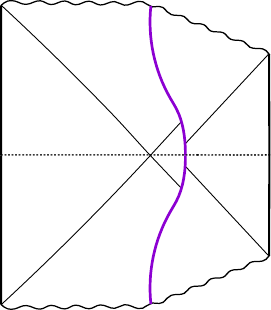} \\
    (a) \hspace{7cm} (b)\hspace{3cm} (c)
    \caption{({\bf a}) The trajectory (purple line) of the massive thin shell in Euclidean AdS$_3$, governed by~\eqref{eq: shell-outside-minus}-\eqref{eq: shell-outside-plus}. The crosses denote horizons $r_-$ and $r_+$. The ``$+$'' and ``$-$'' regions are glued along the shell trajectory. The dashed curves outline the excised part of the spacetime. ({\bf b}) The $r_+$ horizon now belongs to the excised regions such that at the $t=0$ slice, the shell is located behind the horizon $r_-$ but outside the $r_+$ horizon in the Lorentzian geometry. ({\bf c}) From a fully Lorentzian picture, the massive thin shell is emitted from the white hole and turns around at the $t=0$ slice after which it falls back in and is reabsorbed by the black hole. } 
    \label{fig:shelloutside}
\end{figure}
Now consider the shell that is outside the $r_+$ horizon and inside the $r_-$ horizon. At the turning point, we impose the boundary conditions $\tau_\pm(R_*) = 0$. Integrating~\eqref{eq: dtau} gives the elapsed time
\bea
\tau_-(R) &=& \int_{R_*}^{R}  \frac{d\Tilde{R}}{f_-} \sqrt{\frac{f_- + V_{\text{eff}}}{-V_{\text{eff}}}}\,, 
\label{eq: shell-outside-minus}\\
\tau_+(R) &=& \int_{R_*}^{R}  \frac{d\Tilde{R}}{f_+} \sqrt{\frac{f_+ + V_{\text{eff}}}{-V_{\text{eff}}}}\,.\label{eq: shell-outside-plus}
\eea
The spacetime geometry for $d=2$ corresponding to this case is shown in \autoref{fig:shelloutside}. Now, the trajectories are given by
\bea
\tau_-(R) &=& \frac{\beta_-}{2\pi}\arctan\left(\frac{r_- \sqrt{R^2-R_*^2}}{R\sqrt{R_*^2-r_-^2}}\right)\,, \quad (\text{for } d=2)\,, 
\label{eq: shell-outside-minus-d2}\\
\tau_+(R) &=&  \frac{\beta_+}{2\pi}\arctan\left(\frac{r_+ \sqrt{R^2-R^2_*}}{R\sqrt{R_*^2-r_+^2}}\right)\,,\quad (\text{for } d=2) \,.\label{eq: shell-outside-plus-d2}
\eea
The boundaries of the ``$+$"  and ``$-$" geometries have lengths 
\be
\Bar{\beta}_+ = \Delta \tau_+,\quad \beta_- = \Delta \tau_- + \Bar{\beta}_-,
\ee
where $\Delta \tau_\pm$ are given in eq.~\eqref{DeltaTau}. In this case, 
$\sgn(\dot{\tau_\pm}) > 0$ for both components of the spacetime geometry for $d=2$. Eq.~\eqref{eq: sign-dtau} then implies that $\Gamma(R)|_{d=2} > 0$. This implies that the solution with the shell inside the $r_-$ horizon and \textbf{outside} the $r_+$ horizon is consistent when the mass of the shell is \textbf{lower} than $m_0$. 

\paragraph{Lorentzian continuation} 
So far, we have established that for any given value of the shell mass $m$ there is only one saddle point of Euclidean gravity path integral. These solutions can be analytically continued to Lorentzian geometries using the procedure described in~\autoref{app: D=3}. The Hartle-Hawking preparation and the Lorentzian time evolution of these states are pictured in~\autoref{fig:shellinside}(b) and \autoref{fig:shelloutside}(b).
Depending on the value of this mass, in the Lorentzian analytic continuation of the solution the shell will be either inside the horizon or will be outside the horizon at $t_\pm=0$.\footnote{Naively one could also expect a horizonless solution, where both horizons are excised from the Euclidean geometry. However, such a solution would be inconsistent with eq.~\eqref{eq: sign-dtau} in both regions of spacetime for any (positive) value of mass.}

In what follows, we refer to a shell inserted behind both black hole horizons in the semiclassical geometry of the microstate as an \textit{internal shell}. In contrast, a shell inserted behind one of the horizons but outside the other horizon at the $t=0$ slice in the semiclassical geometry of the microstate will be called an \textit{emitted shell}, since it is emitted in the past and reabsorbed in the future, see~\autoref{fig:shelloutside}(c). 

Notice that although we focused on $d=2$ to provide an explicit example, the above arguments can be generalized to higher dimensions. In particular, one considers the value of $\Gamma(R=R_*)$ at the $t_\pm=0$ slice to distinguish whether the shell corresponds to \textit{internal} $(\Gamma(R_*)<0)$ or \textit{emitted}  $(\Gamma(R_*)>0)$.

\subsection{Building blocks of on-shell actions}\label{sec: building blocks}

We now review how to construct the on-shell actions of the geometries dual to the states we have been building. As outlined in the introduction, we take the bulk action to be that of general relativity in the presence of a spherical shell of matter
\begin{equation} \label{eq: I total}
    I=I_{\rm GR} +I_{\rm shell} \,,
\end{equation}
where
\begin{equation}\label{eq: GR and shell}
    I_{\rm GR} = -\frac{1}{16\pi G} \int_{\cal M} \left(R-2\Lambda\right) +\frac{1}{8\pi G}\int_{\partial \cal M} K\,, \quad  I_{\rm shell}=  \int_{\cal W} \sigma\,.
\end{equation}
In the above equations, the bulk manifold is denoted ${\cal M}$, its asymptotic boundary $\partial {\cal M}$, and the shell worldvolume ${\cal W}$. The Gibbons-Hawking-York boundary term ensures that the asymptotic boundary satisfies Dirichlet boundary conditions. Following the holographic renormalization scheme~\cite{Henningson:1998gx,Balasubramanian:1999re,Skenderis:2002wp}, we regulate the action by evaluating it up to a cutoff surface $\partial{\cal M}_\epsilon$ that approaches $\partial \cal M$ as $\epsilon\to 0$ and add the appropriate counterterm $I_{\rm ct}$ such that the total regulated action 
\begin{equation}
    I_{\rm tot}^{\rm reg} = I_{\rm GR}^{\rm reg} + I_{\rm shell}^{\rm reg} + I_{\rm ct}
\end{equation}
is finite in the limit $\epsilon\to 0$. The renormalized action is then given by
\begin{equation} \label{eq: I total ren}
    I^{\rm ren}_{\rm tot}=\lim_{\epsilon\to 0}I^{\rm reg}_{\rm tot}\,.
\end{equation}

On-shell, the bulk equations of motion imply
\begin{equation}
    R-2\Lambda = -2d - \frac{16 \pi G}{d-1} S\,,
\end{equation}
where $S$ is the trace of the matter stress tensor $S_{ab} = -\sigma u_a u_b$, which vanishes everywhere except at the shell worldvolume ${\cal W}$. The on-shell geometries are the black hole solutions in eq.~\eqref{eq: blackholemetric}, for which the extrinsic curvature at the asymptotic boundary is $K=d$. The on-shell action of the shell of matter is an integral of a radially dependent surface density $\sigma$~\eqref{eq: sigma}, and the integral over the worldvolume ${\cal W}$ gives, after integrating the spherical directions, the length of the trajectories $\gamma_{\cal W}$ of the constituent particles 
\begin{equation}
    I_{\rm shell} = m L\left[ \gamma_{\cal W} \right]\,.
\end{equation}

Plugging this back into the action, we get
\begin{equation}
\begin{aligned}
    I^{\rm reg}_{\rm tot}\Big|_{\rm on-shell} &= I^{\rm reg}_{\cal M}+ I^{\rm reg}_{\cal W} + I^{\rm reg}_{\partial \cal M} + I_{\rm ct}\,,
\end{aligned}
\end{equation}
where 
\begin{equation}
    I^{\rm reg}_{\cal M} = \frac{d}{8\pi G} {\rm Vol}\left[ {\cal M}\right]\,, \quad I^{\rm reg}_{\cal W} = \frac{d-2}{d-1} m L\left[\gamma_{\cal W} \right]\,, \quad  I^{\rm reg}_{\partial \cal M} = \frac{d}{8\pi G} {\rm Vol}\left[\partial{\cal M}\right] \,.
\end{equation}
 We have combined the localized contribution of the on-shell Einstein-Hilbert action and the shell action into the on-shell action $I^{\rm reg}_{\cal W}$ evaluated on the worldvolume.

 The (regulated) length of the trajectories is given by
\begin{align}
    L[\gamma_{\cal W}] & =  \int_{-T_{\epsilon}}^{T_{\epsilon}}  dT = 2 \int_{R_*}^{R_{\epsilon}} \frac{dR}{\sqrt{-V_{\rm eff}(R)}} \,,\\
     & = 2\,\text{arccosh}\left(\frac{R_\epsilon}{R_*}\right)\,,\quad (\text{for }d=2)\,,
\end{align}
where $T_\epsilon$ and $R_\epsilon$ are the values of $T$ and $R$ for which the trajectories reach the cutoff surface $\partial {\cal M}_\epsilon$. Notice that the regulated length diverges as $R_\epsilon \xrightarrow{\epsilon\to 0} \infty$. These divergences are canceled by the counterterms, resulting in a finite renormalized length such that
\begin{equation}\label{eq: IW}
    I^{\rm ren}_{\cal W} = \frac{d-2}{d-1} m L^{\rm ren}\left[\gamma_{\cal W} \right]\,.
\end{equation}
As an example, we can consider $d=2$ and find that the renormalized length is given by
\begin{equation}
 L^{\rm ren}\left[\gamma_{\cal W} \right]=2\log\left(\frac{2}{R_*}\right) \,.
\end{equation}

To compute the bulk contribution to the on-shell action, we break down the geometry into sections. For example, for the geometry in~\autoref{fig:shellinside}, we divide it into
\begin{equation}
    \vcenter{\hbox{\includegraphics[width= 10 mm]{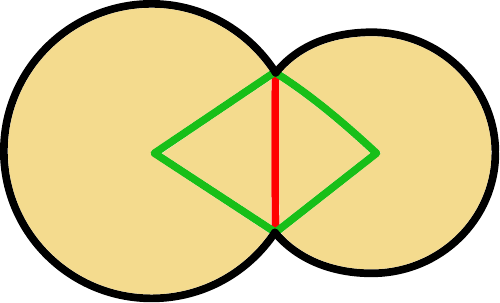}}} = \PL_{\,-}  + \CL_{\,-} + \SR + \CR_{\,+}  + \PR_{\,+} \,.
\end{equation}
The vertical red line indicates the shell worldvolume term $I^{\rm ren}_{\cal W}$ computed in~\eqref{eq: IW}. The remaining contribution $I^{\rm reg}_{\cal M}+ I^{\rm reg}_{\partial {\cal M}}$ is given by
\begin{equation}\label{eq: I contributions}
    I^{\rm reg}_{\cal M} + I^{\rm reg}_{\partial {\cal M}} =  \IPL{\,-}^{\rm reg} + \ICL{\,-}^{\rm reg} + \ICR{\,+}^{\rm reg} + \IPR{\,+}^{\rm reg}\,.
\end{equation}
After including the relevant counterterms from $I_{\rm ct}$, the contributions containing the asymptotic boundaries are 
\begin{equation}\label{eq: I free energy}
    \IPL{\,-}^{\rm ren} = \Bar{\beta}_- {\cal F} (\beta_-)\,, \quad \IPR{\,+}^{\rm ren} = \Bar{\beta}_+ {\cal F} (\beta_+)\,.
\end{equation}
These contributions to the on-shell action are proportional to the thermal free energies ${\cal F}(\beta_\pm)$, which follows from the fact that the part of geometry in which the on-shell action is being evaluated is a fraction of the Euclidean black hole. For holographic CFTs, the free energies are given by~\cite{Emparan:1999pm}
\be\label{eq:renormF}
{\cal F}(\beta_\pm) = \dfrac{V_\Omega}{16\pi G}\left(-r_\pm^{d}+ r_\pm^{d-2} + c_d\right)\;,
\ee
where the constant $c_d$ accounts for the Casimir energy of the CFT in even dimensions \cite{Balasubramanian:1999re} ($c_d = -1,\frac{3}{4}, -\frac{5}{8},\ldots$ in $d=2,4,6,\ldots$)\footnote{The published version of the article is off by a factor of 2. We have corrected the factor to agree with the original references~\cite{Balasubramanian:1999re,Emparan:1999pm} and corrected the intermediate expressions in eqs.~\eqref{eq: free energy d=2} and \eqref{eq: entropy}. All other results are unaffected by the correction.}.
Recall that the horizon radius $r_\pm$ is related to the inverse temperature $\beta_\pm$ by $\beta_\pm = \frac{4\pi}{f_\pm'(r_\pm)}$, where the blackening factor $f_\pm(r)$ is given in~\eqref{eq: fpm}. As an explicit example, for $d=2$
\begin{equation}\label{eq: free energy d=2}
    {\cal F}(\beta_\pm) =- \frac{\pi^2}{2G\beta_\pm^2}  \,, \quad ({\rm for }\ d=2)\,.
\end{equation}

The remaining contributions are proportional to the volumes of the corresponding parts of the geometry
\begin{equation}\label{eq: I vols}
    \ICL{\,-}^{\rm reg} = \frac{d}{8\pi G} {\rm Vol}\left[\CL_{\,-} \right] \,, \quad \ICR{\,+}^{\rm reg} = \frac{d}{8\pi G} {\rm Vol}\left[\CR_{\,+} \right]\,.
\end{equation}
To compute these (regulated) volumes, we integrate the volume element from $r_\pm$ to the location of the shell and then integrate over the trajectory of the shell
\begin{align}\label{eq: volume shell internal minus}
    {\rm Vol}\left[\CL_{\,-} \right] & =\frac{ V_{\Omega}}{d}\int_{-\frac{\Delta\tau}{2}} ^{\frac{\Delta\tau}{2}} \int_{r_-} ^{R(\tau_-)} r^{d-1}   drd\tau_-\nonumber\\
    &= \frac{V_\Omega}{d} \int_{-\frac{\Delta\tau}{2}} ^{\frac{\Delta\tau}{2}}  (R(\tau_-)^d-r_-^d)d\tau_-
    \nonumber\\ & =\frac{2 V_{\Omega}}{d}  \int_{R_*} ^{R_\epsilon}\frac{1}{f_-(R)}\sqrt{\frac{f_-(R)+V_{\rm eff}(R)}{-V_{\rm eff}(R)}}\left(R^d-r_-^d\right)dR\\
    &=2 \pi\sqrt{R_*^2-r_-^2}\, {\rm arccosh}\left( \frac{R_\epsilon}{R_*}\right) \,,\quad (\text{for }d=2)\,,\label{eq: volume shell internal minus d=2}
\end{align}
where in the last step we have used~\eqref{eq: dtau}. Similarly, we have
\begin{align}\label{eq: volume shell internal plus}
    {\rm Vol}\left[\CR_{\,+} \right] &=\frac{2 V_{\Omega}}{d}  \int_{R_*} ^{R_\epsilon}\frac{1}{f_+(R)}\sqrt{\frac{f_+(R)+V_{\rm eff}(R)}{-V_{\rm eff}(R)}}\left(R^d-r_+^d\right)dR\\
    &=2 \pi\sqrt{R_*^2-r_+^2} \, {\rm arccosh}\left( \frac{R_\epsilon}{R_*}\right) \,,\quad (\text{for }d=2)\,,\label{eq: volume shell internal plus d=2}
\end{align}
Following the holographic renormalization prescription, these volumes are renormalized after including the relevant counterterms, leading to
\begin{equation}\label{eq: I vols ren}
    \ICL{\,-}^{\rm ren} = \frac{d}{8\pi G} {\rm Vol}^{\rm ren}\left[\CL_{\,-} \right] \,, \quad \ICR{\,+}^{\rm ren} = \frac{d}{8\pi G} {\rm Vol}^{\rm ren}\left[\CR_{\,+} \right]\,.
\end{equation}
We consider once again the case $d=2$ as an example where we find that
\begin{equation}\label{eq: vols ren}
    {\rm Vol}^{\rm ren}\left[\CL_{\,-} \right]=2\pi\sqrt{R_*^2-r_-^2} \log\left(\frac{2}{R_*}\right) \,, \quad  {\rm Vol}^{\rm ren}\left[\CR_{\,+} \right]=2\pi\sqrt{R_*^2-r_+^2} \log\left(\frac{2}{R_*}\right)\,.
\end{equation}

One can proceed analogously for the emitted shell and break down the geometry in~\autoref{fig:shelloutside} as 
\begin{equation}
    \vcenter{\hbox{\includegraphics[width= 10 mm]{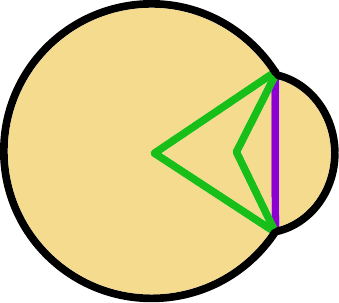}}} = \PL_{\,-} + \CL_{\,-} + \SP - \CLB_{\,+} + \PiR_{\,+}\,,
\end{equation}
and $I^{\rm reg}_{\cal M}+ I^{\rm reg}_{\partial \cal M}$ is given by
\begin{equation}
    I^{\rm reg}_{\cal M} + I^{\rm reg}_{\partial {\cal M}} =  \IPL{\,-}^{\rm reg} + \ICL{\,-}^{\rm reg} - \ICLB{\,+}^{\rm reg} + \IPiR{\,+}^{\rm reg}\,.
\end{equation}
The contributions containing the asymptotic boundaries are the same as in~\eqref{eq: I free energy}, where in particular $\IPiR{\,+}^{\rm reg}$ is functionally the same expression as $\IPR{\,+}^{\rm ren}$. The contributions proportional to the (regulated) volumes of the remaining parts of the geometry are similar to~\eqref{eq: I vols} and are given by
\begin{equation}
    \ICL{\,-}^{\rm reg} = \frac{d}{8\pi G} {\rm Vol}\left[\CL_{\,-} \right] \,, \quad \ICLB{\,+}^{\rm reg} = \frac{d}{8\pi G} {\rm Vol}\left[\CLB_{\,+}\right]\,,
\end{equation}
where ${\rm Vol}\left[\CL_{-} \right]$ is given by~\eqref{eq: volume shell internal minus}, and ${\rm Vol}\left[\CLB_{\,+} \right]$ is the same as~\eqref{eq: volume shell internal plus},
\begin{align}\label{eq: volume shell external plus}
    {\rm Vol}\left[\CLB_{\,+} \right] &=\frac{2 V_{\Omega}}{d}  \int_{R_*} ^{R_\epsilon}\frac{1}{f_+(R)}\sqrt{\frac{f_+(R)+V_{\rm eff}(R)}{-V_{\rm eff}(R)}}\left(R^d-r_+^d\right)dR\,,\\
    &=2\pi\sqrt{R_*^2-r_+^2} \, {\rm arccosh}\left( \frac{R_\epsilon}{R_*}\right) \,.\quad (\text{for }d=2)\,.\label{eq: volume shell external minus d=2}
\end{align}
These contributions will be renormalized by counterterms, similarly to~\eqref{eq: I vols ren} and \eqref{eq: vols ren}.

To conclude, we have provided all the building blocks to compute the on-shell action associated with the geometries shown in~\autoref{fig:shellinside} and \autoref{fig:shelloutside}. Moreover, these building blocks will be useful in computing the on-shell actions of more general geometries, which will be needed in the later sections. For instance, the normalization constant $Z_\Psi$ in \eqref{state with one shell} can be computed as follows 
\begin{equation}\label{eq: norm one shell grav path int}
    \overline{Z}_\Psi = e^{-I_{\rm on-shell}}\,,
\end{equation}
to leading order in $e^{-1/G_N}$. The overbar indicates a computation using the semiclassical saddle points of the effective low energy path integral, which we understand as averaging over the UV degrees of freedom of a complete quantum gravity theory. The resulting approximation is given by the on-shell action of the low energy effective theory.

\subsection{Microcanonical states}
\label{Microcanonical states}

In this section, we set up the microcanonical projection of general states. This microcanonical projection was implicitly used in \cite{Penington19,shells1}, and is explained in more detail in \cite{Climent24,Balasubramanian:2024yxk} in application to one-shell states with equal left and right preparation temperatures. See also~\cite{Marolf:2018ldl} for a related study of microcanonical states in holography. We start by assuming that the Hilbert space is decomposed into microcanonical windows,
\be\label{eq: Hbulk  factors}
\mH = \bigoplus_{E} \mH^{E}\,.
\ee
We use the shorthand notation $E=(E_L,E_R)$ to label each microcanonical band. Each $\mH^E = \mH^{E_L}_L \otimes \mH^{E_R}_R$ factorizes and is spanned by eigenstates of the system with energies in the range $[E_{L, R}, E_{L, R} + \delta E)$. The width $\delta E$  of the band is assumed to be small compared to the temperature scale $\Bar{\beta}^{-1}_{L\,,R}$, but large compared to the average energy spacing $\epsilon$
\begin{equation}
     \epsilon \ll \delta E\ll \frac{1}{\Bar{\beta}_{L\,,R}}\,. \label{eq: limits on width}
\end{equation}
We define the microcanonical projector $\Pi_E = \Pi^L_{E_L} \otimes \Pi^R_{E_R}$ onto the microcanonical window $\mH^E$
\begin{equation}
    \Pi_E: \bigoplus_{E'} \mH^{E'} \to \mH^E.
\end{equation}

To understand the action of this projector, let us consider a general state $|\Phi \rangle \in \mH $ given by
\be
|\Phi \rangle = \frac{1}{\sqrt{{\cal N}_\Phi } } \sum_{n, m} \Phi_{mn} |m\rangle_L \otimes |n\rangle_R\,,
\ee
where the sum goes over the entire spectrum of the doubled boundary system, the normalizing factor is ${\cal N}_\Phi  = \sum_{nm} |\Phi_{nm}|^2$ and the matrix elements $\Phi_{mn}$ include a combination of operator insertions and finite-length Euclidean time evolutions. The microcanonical projector acts on this state as
\be\label{eq: projection}
|\Phi^E \rangle = \sqrt{\frac{\mathcal{N}_\Phi  }{\mathcal{N}_\Phi^E }} \Pi_E |\Phi \rangle\,, \qquad {\cal N}_\Phi  ^E  \equiv {\cal N}_\Phi   \langle \Phi | \Pi_E | \Phi \rangle\,,
\ee
and the microcanonically projected state is supported within the microcanonical window $\mH^E$
\be
|\Phi^E \rangle = \frac{1}{\sqrt{{\cal N}_\Phi^E }} \sum_{|m\rangle_L \otimes |n\rangle_R \in \mH^E } \Phi_{mn} |m\rangle_L \otimes |n\rangle_R\,.
\ee

For the one-shell state $|\Psi \rangle$ in~\eqref{state with one shell}, we have
\be
|\Psi \rangle = \frac{1}{\sqrt{{\cal N}_\Psi } } \sum_{n, m} \Psi_{mn} |m\rangle_L \otimes |n\rangle_R\,,\quad 
\Psi_{mn} = e^{-\frac12 \Bar{\beta}_L E_m - \frac12 \Bar{\beta}_R E_n} \mO_{mn}\,. \label{eq: Psi_mn one shell}
\ee
Since $\delta E \ll \Bar{\beta}^{-1}_{L, R}$, we can write the projected state as
\be\label{eq: microcanonical state}
|\Psi^E \rangle = \frac{1}{\sqrt{\bZ_\Psi}}\sum_{|m\rangle_L \otimes |n\rangle_R \in \mH^E } \mO_{mn} |m\rangle_L \otimes |n\rangle_R\,,
\ee
where $\bZ_\Psi ={\rm Tr}_{{\cal H}}(\Pi_{E_L} \mO\Pi_{E_R}\mO^\dagger) $. 
This normalization constant $\bZ_\Psi$ is related to the normalization of the canonical states $Z_\Psi$ by an inverse Laplace transform\footnote{Note that in this inverse Laplace transform, there are many configurations for which the shell is external on either side of the black hole. This is due to the integral over $\Bar{\beta}_{L,R}$.}
\begin{equation}
    \bZ_\Psi = {\rm Tr}_{{\cal H}}(\Pi_{E_L} \mO\Pi_{E_R}\mO^\dagger) = \frac{\mathbf{h}}{2\pi } \int  d\bar{\beta}_Ld\bar{\beta}_R \, e^{\bar{\beta}_L E_L + \bar{\beta}_R E_R} Z_\Psi\,,\label{eq: bZ1}
\end{equation}
whenever $\delta E \ll \Bar{\beta}_{L,R}^{-1}$.
The integration is performed along the complex contour going from $-i\infty$ to $+i\infty$ remaining to the right of any singularities. The Hessian determinant $\mathbf{h}$ of $-\log Z_\Psi$ is evaluated with respect to the left and right Euclidean lengths
\begin{equation}\label{eq: Hessian}
    \mathbf{h}= -{\rm det} \left[ \partial_A\partial_B \log Z_\Psi\right]\Big|_{\bar{\beta}_{L,R}^*}\,, \quad  {A, B\in \{\bar{\beta}_L,\bar{\beta}_R\}}\,,
\end{equation}
evaluated on the saddle point values $\bar{\beta}^*_{L,R}$. Recall that $\overline{Z_\Psi}$ can be evaluated using a saddle point approximation of the gravity path integral~\eqref{eq: norm one shell grav path int}. In this case, $-\log \overline{Z_\Psi}= I_{\rm on-shell}$ and the saddle point values $\bar{\beta}_{L,R}^*$ are found by
\begin{equation}\label{eq: saddle point}
    \partial_{\bar{\beta}_{L,R}} I_{\rm on-shell}\Big|_{\bar{\beta}^*_{L,R}} = E_{L,R}\,. 
\end{equation}

We are now fully equipped to compute the overlap statistics of the semiclassical black hole microstate basis for black holes in the canonical and microcanonical ensembles, which we will extensively use in the following.

\section{How to count states}
\label{sec:review}

In this section, we summarize the results in~\cite{shells1} and clarify some subtleties that were previously omitted. We can leverage the framework described in~\autoref{sec:setup} to construct an infinite family of states $| \Psi_k \rangle$ of the form~\eqref{state with one shell}, all of which are dual to an eternal AdS black hole geometry outside the horizon but which differ inside the horizon. This infinite family forms an overcomplete basis of black hole microstates. We begin with the definition of these states in a double copy of a CFT  
\be\label{eq: single-shell microstates}
| \Psi_k \rangle = \frac{1}{\sqrt{Z_{kk}}} \sum_{n, m} e^{-\frac12 \Bar{\beta}_L E_n - \frac12 \Bar{\beta}_R E_m} \mO^{(k)}_{nm} |n \rangle_L \otimes |m \rangle_R\,,
\ee
where $Z_{kk} = {\rm Tr} \left[{\cal O}^{(k)\dagger} e^{-\Bar{\beta}_L H} {\cal O}^{(k)}e^{-\Bar{\beta}_R H}\right]$ normalizes the states. The operators ${\cal O}^{(k)}$ are products of $n$ scalar operators ${\cal O}_{\Delta_k}$ with conformal dimension $\Delta_k$ that are uniformly supported on the $(d-1)$-dimensional spherical time slice of the CFT. As explained in \autoref{sec: definition}, the bulk dual to these states corresponds to two asymptotically AdS geometries connected by a wormhole. The operator ${\cal O}^{(k)}$ sources a spherical shell of matter whose backreaction elongates the wormhole. The mass of the shell of matter is given by 
\be\label{eq: mass Ok}
m_k = k \mu\,, \qquad k = 1,\ 2,\dots,\ \Omega,
\ee
and the dimension $\Delta_k$ of the operator ${\cal O}_{\Delta_k}$ satisfies $\mu^2k^2 = n^2 \Delta_k (\Delta_k - d)$.

As explained in \autoref{Microcanonical states}, we can project these states onto the microcanonical band $[E_{L,\, R},\, E_{L, \,R} + \delta E)$ using a projector $\Pi_E = \Pi_E^L \otimes \Pi_E^R$,
\be
| \Psi^E_k \rangle =\frac{1}{\sqrt{\langle \Psi_k|\Pi_E|\Psi_k\rangle}}\, \Pi_E | \Psi_k \rangle \,.\label{micro-state}
\ee
We are interested in computing the dimension of the space spanned by $\Omega$ of these normalized states $| \Psi^E_k \rangle$ 
\begin{equation}\label{eq:Kspanmicro}
    {\cal H}^E_{\rm bulk}(\Omega) \equiv {\rm Span}\{|\Psi^E_k\rangle, \quad k=1,\ldots,\Omega\}\,.
\end{equation}
From the boundary point of view, the dimension of ${\cal H}^E_{\rm bulk}(\Omega)$ is upper bounded by $\rm{dim}\left(\mathcal{H}_L\right)\times \rm{dim}\left(\mathcal{H}_R\right)$. This can be argued by counting the number of independent coefficients in the matrix $ \mO^{(k)} _{mn}$, considering that the index $m$ is restricted to run over $\rm{dim}\left(\mathcal{H}_L\right)$ different states and similarly the index $n$ runs over $\rm{dim}\left(\mathcal{H}_R\right)$  different states. When this matrix is random, consistent with the eigenstate thermalization hypothesis (ETH) \cite{Srednicki1994}, the dimension of the Hilbert space will saturate this bound. On the other hand, as will be shown below, from the bulk gravity perspective, the states in~\eqref{eq:Kspanmicro} are orthogonal to leading order in the semiclassical approximation. From the gravity perspective, therefore, it would seem like the dimension of ${\cal H}^E_{\rm bulk}\left( \Omega\right)$  would increase without bound as $\Omega$ increases. With the above discussion in mind, the work of~\cite{shells1,shells2} is a purely gravitational computation which shows that, because of quantum gravity effects, and so long as  $\Omega$ is large enough, this set of states in fact spans a Hilbert space with a dimension precisely equal to the exponential of the Bekenstein-Hawking entropy of the black hole horizons and hence saturates the bound. In the rest of this section, we provide a pedagogical summary of the results of~\cite{shells1}, which will introduce the necessary framework to derive our main results in~\autoref{sec:mainresult}. Specifically, ref.~\cite{shells1} showed that when $\Omega$ is large enough, the vectors in this family are linearly dependent and the auxiliary Hilbert space ${\cal H}_{\rm bulk}^E (\Omega)$ is the true microcanonical Hilbert space ${\cal H}_{\rm bulk}^E$ with dimension $\rm{dim}\left(\mathcal{H}_L\right)\times \rm{dim}\left(\mathcal{H}_R\right)$. To diagnose this effect, we introduce the Gram matrix of the microstate overlaps,
\be
G_{ij} = \langle \Psi^E_i | \Psi^E_j \rangle\,.
\ee
The aim is to study the eigenvalues of the Gram matrix while varying $\Omega$. The overlaps are computed using the saddle point approximation of the gravitational path integral, which we denote by an overline. Specifically, to leading order in the semiclassical approximation, the states $|\Psi^E_i \rangle$ and $|\Psi^E_j \rangle$ are orthogonal, i.e., $\overline{\langle \Psi^E_i|\Psi^E_j \rangle} \propto \delta_{ij}$. However, wormhole contributions to the gravitational path integral induce a small overlap 
\cite{shells1,shells2,Climent24}. This non-orthogonality leads to saturation of the dimension of~\eqref{eq:Kspanmicro} to precisely the exponential of the Bekenstein-Hawking entropy \cite{shells1,shells2,Climent24}. Following \cite{shells1}, the main steps are the following: 
\begin{itemize}
    \item[1.] Using the tools of~\autoref{sec: building blocks}, obtain the expression for the moments of the overlaps evaluated with the gravity path integral (denoted by the overline) 
    \be\label{eq: moments of overlaps]}
 \overline{\langle \Psi_i | \Psi_{k_1}  \rangle \langle \Psi_{k_1} | \Psi_{k_2}  \rangle \dots \langle \Psi_{k_{n-1}} | \Psi_j  \rangle}\
    \ee
    for general $n$.
    \item[2.] Perform a microcanonical projection to compute the overlaps of the microcanonical states and sum over the indices $k_i$ to compute
    \be\label{eq: micromoments of overlaps]}
\left(G^n\right)_{ij} = \sum_{k_1,\ldots,k_n}^\Omega \overline{\langle \Psi^E_i | \Psi^E_{k_1}  \rangle \langle \Psi^E_{k_1} | \Psi^E_{k_2}  \rangle \dots \langle \Psi^E_{k_{n-1}} | \Psi_j^E \rangle}\,.
    \ee
    \item[3.] Define the resolvent  of the Gram matrix $G$
\be
{R_{ij}(\lambda)} := {\left(\frac{1}{\lambda \mathds{1}    - G}\right)_{ij}} = \frac{1}{\lambda}\delta_{ij} + \sum_{n=1}^\infty \frac{1}{\lambda^{n+1}} {(G^n)_{ij}}\,. \label{eq: resolvent}
\ee
The trace of the resolvent $R(\lambda) = \sum_{i=1}^\Omega {R_{ii}(\lambda)}$ will have poles at each eigenvalue of $G$, and the residue of each pole counts the degeneracy of the corresponding eigenvalue. Of particular interest is the value of $\Omega$ at which $G$ first develops zero eigenvalues, which can be examined using the asymptotic behavior of 
 $R$ as $\lambda\to 0$. Further increasing the value of $\Omega$ increases only the number of null states in~\eqref{eq:Kspanmicro}. Since the number of linearly independent states remains unchanged, the value of $\Omega$ at which null states emerge corresponds to the dimension of the microcanonical Hilbert space.

\end{itemize}

In what follows, we will work to leading order in \( G_N \), \( \Omega^{-1} \), and \( e^{-1/G_N} \). The Newton's constant $G_N$ is assumed to be small, ensuring the validity of the semiclassical approximation to the gravitational path integral. The convergence of the auxiliary Hilbert space ${\cal H}_{\rm bulk}^E (\Omega)$ to the true microcanonical Hilbert space ${\cal H}_{\rm bulk}^E$ happens when  $\Omega \gtrsim e^{1/G_N}$~\cite{shells1}, so $\Omega$ is assumed to be large. To simplify the computations, we will also restrict to states in the set~\eqref{eq:Kspanmicro} whose gravitational duals include shells with very large mass. To this end, we provide results to leading order in large $\mu$.

\subsection{Moments of overlaps}
\label{sec: Overlaps}
The moments of overlaps~\eqref{eq: moments of overlaps]} are the key ingredient needed to compute the dimension of the Hilbert space~\eqref{eq:Kspanmicro}, and we will use the gravity path integral to compute them. This means that the CFT path integral contours which prepare each of the $n$ copies of states $\Psi_j$ and $\Psi_j$ serve as boundary conditions for the dual AdS spacetime geometries. At the leading semiclassical level, computing the gravity path integral amounts to computing the on-shell action of all solutions of the Einstein equations with a negative cosmological constant which satisfy these boundary conditions, and summing over their contributions. 

\paragraph{Inner product ${\langle \Psi_i | \Psi_j \rangle}$}
\begin{figure}[t]
    \centering
    \includegraphics[scale=0.60]{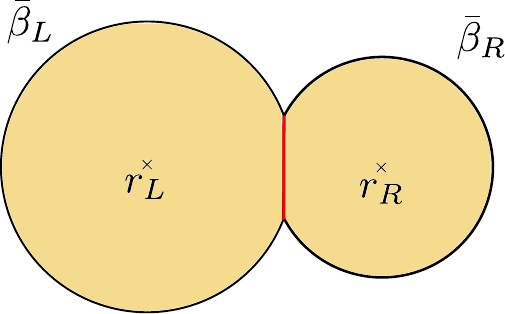} \hspace{1cm}
    \includegraphics[scale=0.60]{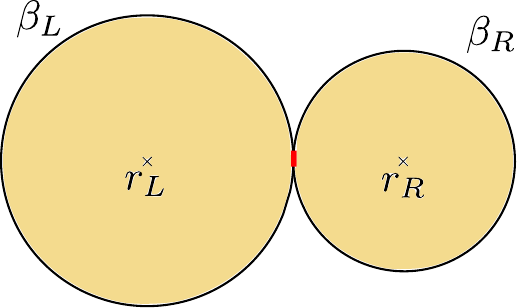} \\
    (a) \hspace{7cm} (b)
    \caption{(a) The geometry computing the overlaps $\overline{\langle \Psi_i | \Psi_i \rangle}$ with a shell inside the black hole horizons $r_L$ and $r_R$. The vertical red line shows the shell worldvolume. (b) The large shell mass limit of the same geometry. The shell worldvolume pinches off into a dot, which turns the yellow regions into full disks. In this limit, $\Bar{\beta}_{L,R}\to \beta_{L,R}$.} 
    \label{fig:internalshell}
\end{figure}
The inner product of states with shells of mass $m_i$ and $m_j$ can be written as
\begin{equation}
    \langle \Psi_i | \Psi_j \rangle = \frac{Z_{ij}}{\sqrt{Z_{ii}Z_{jj}}} \,, \qquad Z_{ij} = {\rm Tr} \left[{\cal O}^{(i)\dagger} e^{-\Bar{\beta}_L H} {\cal O}^{(j)}e^{-\Bar{\beta}_R H}\right]\,.
\end{equation}
We can compute (products of) the normalization coefficients $Z_{ij}$ using the gravity path integral, and in particular, we use the semiclassical approximation, which we denote with an overline. This amounts to finding the classically allowed geometries filling in the boundary conditions specified by (products of) $Z_{ij}$ and computing the on-shell action as explained in~\autoref{sec: building blocks},
\begin{equation}\label{eq: overlap z/zz}
    \overline{\langle \Psi_i | \Psi_j \rangle} = \frac{\overline{Z_{ij}}}{\sqrt{\overline{Z_{ii}} \overline{Z_{jj}}}} \,.
\end{equation}

For states with shells of masses $m_i$ and $m_j$, $Z_{ij}$ is given by the exponential of minus the action of the Euclidean geometry which fills a contour similar to that shown in~\autoref{fig:contouroneshell}(b), but with potentially different operators ${\cal O}^{(i)\dagger}$ and ${\cal O}^{(j)}$. Assuming a finite number of interacting matter fields, no solution of the gravity equations of motion satisfies the boundary conditions given by the contour in~\autoref{fig:contouroneshell}(b) when $m_i \neq m_j$~\cite{shells1}. This means that for $i \neq j$, we have $\overline{\langle \Psi_i | \Psi_j \rangle} = 0$. When the shells are of equal mass, $m_j = m_i = i \mu$, this geometry consists of two segments of the Euclidean AdS black hole of masses $M_L$ and $M_R$, glued to each other along the worldvolume of the shells. When $m_i$ is larger than the critical mass $m_0$ given in~\eqref{eq: m_0}, the geometry is given in~\autoref{fig:internalshell}(a). The on-shell action that computes $\overline{Z_{ii}}$ is given by
\be
I_{\rm tot}^{\rm ren}\Big|_{\rm on-shell}=\IPL{\,L}^{\rm ren}+\ICL{\,L}^{\rm ren}+I_{\textcolor{red}{\cal W}}^{\rm ren} +\ICR{\,R}^{\rm ren} +\IPR{\,R}^{\rm ren}\,,
\ee
where the various contributions to this action are given in \autoref{sec: building blocks}.

In the large $\mu$ limit, $m_i$ becomes large as well, and the worldvolume of the shell becomes infinitesimally short. In this limit, the spacetime geometry pinches off into two Euclidean black holes \cite{shells1}, see~\autoref{fig:internalshell}(b). Evaluating the action leads to
\be\label{eq: norm}
\overline{Z_{ii}} = Z(\beta_L) Z(\beta_R)e^{-I_{\rm univ.}}\,. 
\ee
Here $Z(\beta)$ is the thermal partition function of the Euclidean AdS black hole~\cite{Gibbons1977}, which can be written as 
\be
Z(\beta) = e^{S_{BH} - \beta M} =e^{-\beta\mathcal{F}(\beta)}\,, \label{eq: first law}
\ee
where $S_{BH} = \frac{A}{4 G}$ is the entropy of the black hole with inverse Hawking temperature $\beta$ and $M$ is its mass. The other term,
\begin{equation}
    I_{\rm univ.} = \lim_{R_*\to\infty}\left(\ICL{\,L}^{\rm ren}+I_{\textcolor{red}{\cal W}}^{\rm ren}+\ICR{\,R}^{\rm ren}\right)\,,
\end{equation}
is independent of the details of the geometry, and only depends on the operators ${\cal O}^{(i)}$ and ${\cal O}^{(i)\dagger}$ through the masses $m_i$.

Similarly, we can use the semiclassical approximation to compute the product of normalization factors. For the product $Z_{ii} Z_{jj}$, the leading order contribution is given by two disconnected geometries, where each geometry is depicted in~\autoref{fig:internalshell}. Once again in the large $\mu$ limit, we find
\be\label{eq: ZiiZjj}
\overline{Z_{ii}Z_{jj}} = Z(\beta_L)^2 Z(\beta_R)^2e^{-2I_{\rm univ.}}\,.
\ee
Combining~\eqref{eq: norm} and~\eqref{eq: ZiiZjj} in~\eqref{eq: overlap z/zz}, we find
\be\label{eq: overlap semi}
\overline{\langle \Psi_i | \Psi_j \rangle} =\delta_{ij}\,.
\ee
While we have computed~\eqref{eq: overlap semi} in the large $\mu$ limit, it remains true as long as $\mu$ is finite and non-zero.

\begin{figure}[t]
    \centering
    \includegraphics[scale=0.7]{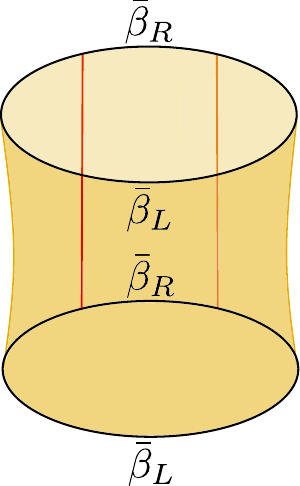} \hspace{1cm} 
    \includegraphics[scale=0.7]{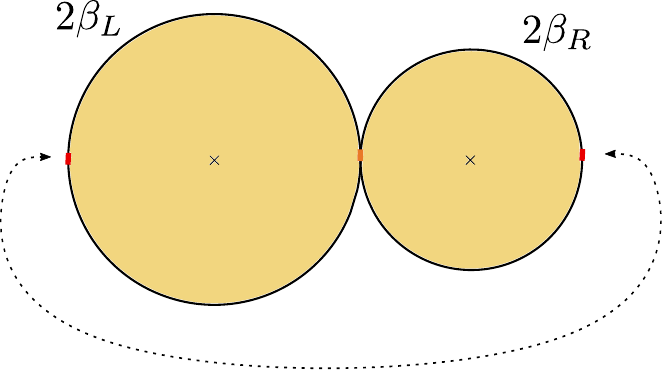} \\
    (a) \hspace{7cm} (b)
    \caption{Connected wormhole geometry contributing to the second moment of the overlap. The red and orange lines are the internal shells of matter of mass $m_i$ and $m_j$ correspondingly. This geometry consists of two segments of the Euclidean AdS black hole spacetime, which are glued together in the cylindrical topology along the shell worldvolumes, as shown in (a). Before taking the pinching limit, the shell trajectories have finite length, and the Euclidean preparation times have lengths $\Bar{\beta}_{L,R}<\beta_{L,R}$. (b) In the pinching limit, the shell trajectories shorten and $\Bar{\beta}_{L,R}\to\beta_{L,R}$. The two Euclidean black hole segments become full disks of circumference $2\beta_L$ and $2\beta_R$, respectively. The red shell is identified on the left and right, indicated by the dotted line.} 
    \label{fig:internalshellwormhole2}
\end{figure}

\paragraph{Higher moments of the overlaps $\langle \Psi_i | \Psi_{k_1} \rangle \dots \langle \Psi_{k_{n-1}} | \Psi_{j} \rangle$} The leading semiclassical result for the overlaps~\eqref{eq: overlap semi} seems to imply that the states are orthogonal. However, the leading result does not capture a small non-perturbative overlap that spoils orthogonality.\footnote{Small corrections to overlaps of this kind can dramatically change the dimensionality of the Hilbert space \cite{shells1} and the nature of the bulk-boundary holographic map \cite{Antonini:2024yif}.} 
This small non-orthogonality is captured by higher moments of the overlaps, which can be computed analogously to the inner product. For example, the second moment of the overlaps is given by
\begin{equation}
    \langle \Psi_i | \Psi_j \rangle \langle \Psi_j | \Psi_i \rangle = \frac{Z_{ij} Z_{ji}}{Z_{ii} Z_{jj}}\,.
\end{equation}
Similarly to the computation of the inner product, we can proceed by taking a semiclassical approximation, which gives
\begin{equation}
    \overline{\langle \Psi_i | \Psi_j \rangle \langle \Psi_j | \Psi_i \rangle} = \frac{\overline{Z_{ij} Z_{ji}}}{\overline{Z_{ii} Z_{jj}}}\,.
\end{equation}
\begin{figure}[t]
    \centering
    \includegraphics[scale=0.6]{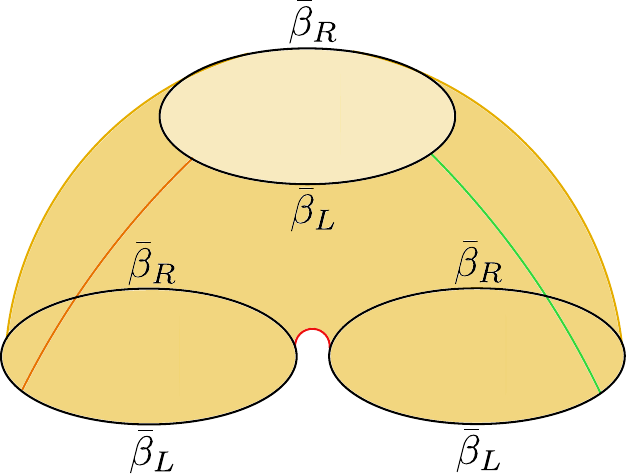} \hspace{1cm} 
    \includegraphics[scale=0.6]{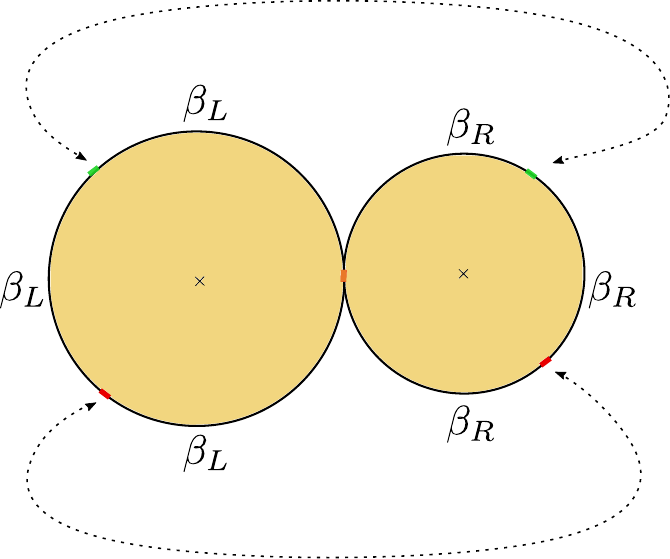} \\
    (a) \hspace{7cm} (b)
    \caption{Three-boundary wormhole which computes the connected contribution to $\overline{Z_{ik}Z_{kj}Z_{ji}}$. The spacetime consists of two segments of Euclidean AdS black hole spacetime, glued to each other along three copies of the shell worldvolume (colored in red, orange and green) in the pair of pants topology as shown in (a) Before the pinching limit, the shell worldvolumes symmetrically connect three asymptotic boundaries, and the Euclidean preparation time segments have lengths $\Bar{\beta}_{L,R}<\beta_{L,R}$. (b) In the pinching limit, the shell worldvolumes contract, $\Bar{\beta}_{L,R}\to \beta_{L,R}$ and the geometry becomes two disks of circumference $3\beta_L$ and $3\beta_R$, respectively. The dashed arrows show identifications along the shell points.} 
    \label{fig:internalshellwormhole3}
\end{figure}

To compute the terms appearing in the second moment of the overlap, one needs to take two copies of the contour in \autoref{fig:contouroneshell}(b) and fill them in with solutions to the gravity equations of motion. If $m_i=m_j$, two geometries contribute, namely a disconnected and a connected one. The former corresponds to two copies of the geometry that computes the norm of the state shown in~\autoref{fig:internalshell}.
The connected contribution corresponds to the two-boundary wormhole depicted in~\autoref{fig:internalshellwormhole2}. The wormhole consists of segments of the Euclidean AdS black hole spacetime, glued along the two copies of the internal shell worldvolume in a cylindrical topology.\footnote{Naively, one can produce a two-boundary wormhole by filling in the boundary condition set by contours in two different ways, which would yield two gravity saddles to sum over. However, as we explain in~\autoref{sec: orientation}, the condition of continuity of Euclidean time in each segment of the Euclidean AdS fixes the orientation of the asymptotic boundaries and makes the solution unique.} Note that the connected wormhole geometry exists as a solution even if $m_i \neq m_j$. The on-shell action that computes this contribution to $\overline{Z_{ij}Z_{ji}}$ is given by
\begin{equation}
    I_{\rm tot}^{\rm ren}\Big|_{\rm on-shell}=2(
    \IPL{\,L}^{\rm ren}+\ICL{\,L}^{\rm ren}+I_{\textcolor{red}{\cal W}}^{\rm ren} +\ICR{\,R}^{\rm ren} +\IPR{\,R}^{\rm ren} )\,.
\end{equation}
Recall that the various contributions to this action are given in \autoref{sec: building blocks} where the temperatures $\beta_{\pm}$ and radii $r_{\pm}$ should be replaced by the temperatures of the wormhole geometry to the left and right of the shells.

The pinching limit of large shell mass $\mu$ shortens both shell worldvolumes, which turns the wormhole geometry into two disks\footnote{Throughout the paper for brevity we use the term ``disk" of given circumference as a substitute for the Euclidean AdS$_{d+1}$ black hole geometry of a given time periodicity.} of circumference $2\beta_{L}$ and $2\beta_R$, respectively.\footnote{Each disk has circumference $2\beta_{L,R}$ because it consists of two asymptotic boundaries with length $\beta_{L,R}$ with approaching endpoints as the shell trajectories shorten in the pinching limit.} In this limit, we find
\begin{equation}\label{eq: ZijZji}
    \overline{Z_{ij}Z_{ji}}\big|_{\rm conn.} = Z(2\beta_L) Z(2\beta_R) e^{-2 I_{\rm univ.}}\,,
\end{equation}
where we recall that the thermal partition is given in eq.~\eqref{eq: first law}.

Combining eqs.~\eqref{eq: ZiiZjj} and~\eqref{eq: ZijZji}, we find the semiclassical approximation to the second moment of the overlaps
\be
\overline{\langle \Psi_i | \Psi_{j} \rangle \langle \Psi_{j} | \Psi_{i} \rangle} = \delta_{ij} + \frac{Z(2\beta_L) Z(2\beta_R)}{Z(\beta_L)^2 Z(\beta_R)^2}\,. \label{eq: overlap-2}
\ee
The Kronecker delta term reflects the fact that the disconnected contribution is present only when $m_i=m_j$, while the connected wormhole geometry exists as a solution even if $m_i \neq m_j$. The result~\eqref{eq: overlap-2} is consistent with the semiclassical approximation of the overlap~\eqref{eq: overlap semi}. The second term captures the size of the off-diagonal corrections that are averaged over in the semiclassical approximation of~\eqref{eq: overlap semi}. These contributions will be of crucial importance for later calculations, so we introduce the notation
\be
\left.\overline{\langle \Psi_i | \Psi_{j} \rangle \langle \Psi_{j} | \Psi_{i} \rangle}\right|_{\text{conn.}} \equiv \overline{\langle \Psi_i | \Psi_{j} \rangle \langle \Psi_{j} | \Psi_{i} \rangle} - \overline{\langle \Psi_i | \Psi_{j} \rangle }\,\overline{\langle \Psi_{j} | \Psi_{i} \rangle} =\frac{Z(2\beta_L) Z(2\beta_R)}{Z(\beta_L)^2 Z(\beta_R)^2}\,. \label{eq: overlap-2 conn}
\ee

A similar computation can be done for generic $n$-th moments of the overlaps
\begin{equation}
    \langle \Psi_{k_1} | \Psi_{k_2} \rangle \langle \Psi_{k_2} | \Psi_{k_3} \rangle \ldots \langle \Psi_{k_{n}} | \Psi_{k_1} \rangle = \frac{ Z_{k_1 k_2}Z_{k_2 k_3}\ldots Z_{k_{n} k_1}}{Z_{k_1 k_1}Z_{k_2 k_2} \ldots Z_{k_{n} k_{n}}}\,,
\end{equation}
which can be computed semiclassically as described above,
\begin{equation}\label{eq: overlaps semiclassical}
    \overline{\langle \Psi_{k_1} | \Psi_{k_2} \rangle \langle \Psi_{k_2} | \Psi_{k_3} \rangle \ldots \langle \Psi_{k_{n}} | \Psi_{k_1} \rangle} = \frac{\overline{ Z_{k_1 k_2}Z_{k_2 k_3}\ldots Z_{k_{n} k_1}}}{\overline{Z_{k_1 k_1}Z_{k_2 k_2} \ldots Z_{k_{n} k_{n}}}}\,.
\end{equation}
We will compute the contribution from the fully connected geometry, namely the $n$-boundary wormhole.\footnote{With the fully connected contributions to all moments of overlaps, one can directly compute the moments of overlaps. Note that this is not needed since later we turn the resolvent equation from containing $G^n$ in eq.~\eqref{eq:resolve-bar} to $G^n_{\rm conn.}$ in eq.~\eqref{eq:resolve-bar-connected}.} This geometry is obtained by gluing $n$ copies of the contour such that in the large mass limit it leads to two disks of circumference $n\beta_L$ and $n\beta_R$. For example, in the case of $n=3$ the connected contribution is given by the pair of pants geometry shown in~\autoref{fig:internalshellwormhole3}. The result associated with the $n$-boundary wormhole is
\begin{equation}\label{eq: Zn conn}
    \overline{ Z_{k_1 k_2}Z_{k_2 k_3}\ldots Z_{k_{n} k_1}}\big|_{\rm conn.} = Z(n\beta_L) Z(n\beta_R) e^{-n I_{\rm univ.}}\,.
\end{equation}
To leading order in the semiclassical approximation, the term in the denominator of eq.~\eqref{eq: overlaps semiclassical} is always dominated by the fully disconnected geometry, giving 
\begin{equation}\label{eq: Zn dis}
    \overline{Z_{k_1 k_1}Z_{k_2 k_2} \ldots Z_{k_{n} k_{n}}} = Z(\beta_L)^n Z(\beta_R)^ne^{-n I_{\rm univ.}}\,.
\end{equation}
Hence, the result for the connected part of the $n$-th moment of the overlaps reads 
\be
\left.\overline{\langle \Psi_{k_1} | \Psi_{k_2} \rangle \langle \Psi_{k_2} | \Psi_{k_3} \rangle \ldots \langle \Psi_{k_{n}} | \Psi_{k_1} \rangle} \right|_{\text{conn.}} = \frac{Z(n\beta_L)Z(n\beta_R)}{ Z(\beta_L)^n Z(\beta_R)^n}\,. \label{eq: overlap-n}
\ee

\subsection{Microcanonical projection}
\label{sec: microoverlaps}

Now that we have computed the moments of overlaps in the canonical states using the gravitational path integral, we compute moments of overlaps for the microcanonical states described in~\autoref{Microcanonical states}. The motivation to work in this basis is twofold. From a technical point of view, while analytic expressions for the overlaps of the canonical states can be found as in~\eqref{eq: overlap-n}, computing the dimension of the Hilbert space spanned by these states is subtle. From a physics perspective, we want to ensure that the Hilbert space spanned by the basis we work with is finite dimensional. Projecting all the states to an energy band ensures that the resulting Hilbert space~\eqref{eq:Kspanmicro} is finite dimensional even as $\Omega$ increases indefinitely. When $\Omega$ becomes large enough, the auxiliary Hilbert space ${\cal H}^E_{\rm bulk}(\Omega)$ becomes the microcanonical Hilbert space ${\cal H}^E_{\rm bulk}$, which is finite dimensional.

The computation of the overlaps in the microcanonical basis is analogous to~\autoref{sec: Overlaps}. From the definition of the microcanonical states~\eqref{eq: microcanonical state}, we find
\begin{equation}\label{eq: overlap microcanonical}
    \langle \Psi_i^E | \Psi_j^E \rangle = \frac{\bZ_{ij}}{\sqrt{\bZ_{ii}\bZ_{jj}}} \,, \qquad \bZ_{ij} = {\rm Tr} \left[ \Pi_{E_L} {\cal O}^{(j)}\Pi_{E_R} {\cal O}^{(i)\dagger}\right] \,.
\end{equation}

The microcanonical normalization factors $\bZ_{ij}$ can be computed from the canonical normalization factors $Z_{ij}$ with an inverse Laplace transform as in eq.~\eqref{eq: bZ1}
\begin{equation}
    \bZ_{ij} = {\rm Tr}_{{\cal H}}(\Pi_{E_L} \mO^{(j)}\Pi_{E_R}\mO^{(i)\dagger}) = \frac{\mathbf{h}_{ij}}{2\pi } \int  d\bar{\beta}_Ld\bar{\beta}_R \, e^{\bar{\beta}_L E_L + \bar{\beta}_R E_R} Z_{ij}\,,\label{eq: bZij}
\end{equation}
where $\mathbf{h}_{ij}$ is the Hessian determinant of $- \log Z_{ij}$ with respect to the inverse temperatures $\bar{\beta}_{L,R}$ evaluated at the saddle point values, see eqs.~\eqref{eq: Hessian} and~\eqref{eq: saddle point}. We can compute the quantities in~\eqref{eq: overlap microcanonical} using the semiclassical approximation of $Z_{ij}$ and the saddle point approximation of the inverse Laplace transform~\eqref{eq: bZij}, which leads to
\begin{equation}
    \overline{\bZ_{ij}} = \delta_{ij} e^{\bS_L+\bS_R-I_{\rm univ.}}\,, \quad {\rm and} \quad \langle \overline{\Psi_i^E | \Psi_j^E \rangle} = \delta_{ij}\,,
\end{equation}
where $\bS_{L,R}$ are equal to the microcanonical entropies counting the number of states in a single copy of the CFT in the respective energy window.

For higher moments of the overlaps, we have
\begin{equation}\label{eq: nth overlap}
    \langle \Psi^E_{k_1} | \Psi^E_{k_2} \rangle \langle \Psi^E_{k_2} | \Psi^E_{k_3} \rangle \ldots \langle \Psi^E_{k_{n}} | \Psi^E_{k_1} \rangle = \frac{ \bZ_{k_1 k_2}\bZ_{k_2 k_3}\ldots \bZ_{k_{n} k_1}}{\bZ_{k_1 k_1}\bZ_{k_2 k_2} \ldots \bZ_{k_{n} k_{n}}}\,,
\end{equation}
which can be computed using a generalization of~\eqref{eq: bZij} 
\begin{equation}
\begin{aligned}
    \bZ_{i_1j_1} \bZ_{i_2j_2}\ldots \bZ_{i_nj_n} & = {\rm Tr}_{{\cal H}}(\Pi_{E_L} \mO^{(j_1)}\Pi_{E_R}\mO^{(i_1)\dagger}) \ldots {\rm Tr}_{{\cal H}}(\Pi_{E_L}\mO^{(j_n)}\Pi_{E_R}\mO^{(i_n)\dagger} ) \\
    & = \frac{\mathbf{H}}{(2\pi)^n } \int  d^n\bar{\beta}_L d^n\bar{\beta}_R \, e^{ \left(\bar{\beta}^{(1)}_L+\ldots +\bar{\beta}^{(n)}_L\right) E_L + \left(\bar{\beta}^{(1)}_R+\ldots +\bar{\beta}^{(n)}_R\right) E_R} Z^{(1)}_{i_1j_1} \ldots Z^{(n)}_{i_nj_n}\,,\label{eq: bZn}
\end{aligned}
\end{equation}
where $d^n\Bar{\beta}_{L,R} = \prod_{i=1}^n d\Bar{\beta}_{L,R}^{(i)}$, and $Z^{(i)}_{ij} = Z_{ij}(\Bar{\beta}_L^{(i)},\Bar{\beta}_R^{(i)})$ and ${\bf H}$ is the Hessian determinant of $-\log  Z^{(1)}_{i_1j_1} \ldots Z^{(n)}_{i_nj_n}$ with respect to $\Bar{\beta}_{L,R}^{(i)}$, evaluated at the saddle point value. 

As a simple example, let us focus on the connected contribution to the $n$-th moment of the overlap~\eqref{eq: nth overlap}. To this end, we compute the inverse Laplace transform of $\overline{ Z_{k_1 k_2}Z_{k_2 k_3}\ldots Z_{k_{n} k_1}}\big|_{\rm conn.}$  with different $\beta_{L,R}^{(i)}$, which in the large $\mu$ limit reduces to a generalization of eq.~\eqref{eq: Zn conn} where $n\beta_{L,R} $ is replaced by $ \sum_i \Bar{\beta}^{i}_{L,R}$ and recall that $\Bar{\beta}^{(i)}_{L,R}={\beta}^{(i)}_{L,R}$ in this limit. In this case, the integrand only depends on the sum of $\Bar{\beta}_{L,R}^{(i)}$. Therefore the $2n$ integrals over $\Bar{\beta}^{(i)}_{L,R}$ reduce to two integrals, which we parametrize by their average $q_{L,R} = \frac{1}{n}\sum_i \Bar{\beta}^{i}_{L,R}$, resulting in
\begin{equation}
\begin{aligned}
    \left.\overline{\bZ_{k_1 k_2}\bZ_{k_2 k_3}\ldots \bZ_{k_{n} k_1}} \right|_{\text{conn.}}
    &=\frac{\mathbf{h}_n}{2\pi } \int  dq_Ldq_R \, e^{n q_L E_L + n q_R E_R} Z(n q_L) Z(n q_R) e^{-n I_{\rm univ.}}\,,\\
    &=e^{\bS_L+\bS_R-nI_{\rm univ.}}.\label{eq: bZn conn num}
\end{aligned}
\end{equation}
Here ${\bf h}_n$ is the Hessian determinant of $-\log Z(n q_L) Z(n q_R) $ with respect to $q_{L,R}$ evaluated at the saddle point. 

The term in the denominator of~\eqref{eq: nth overlap} is dominated by the contribution from the fully disconnected geometry~\eqref{eq: Zn dis}, resulting in
\begin{equation}
\begin{aligned}
    \overline{\bZ_{k_1 k_1}\bZ_{k_2 k_2} \ldots \bZ_{k_{n} k_{n}}}  
    & = \left(\frac{\mathbf{h}_1}{2\pi }\right)^n \int  \prod_{i=1}^n d\bar{\beta}^{(i)}_L d\bar{\beta}^{(i)}_R \, e^{ \bar{\beta}^{(i)}_L E_L+\bar{\beta}^{(i)}_R E_R- I_{\rm univ.}} Z(\bar{\beta}_L^{(i)})Z(\bar{\beta}_R^{(i)})
    \\
    &=e^{n\bS_L+n\bS_R-nI_{\rm univ.}}.\label{eq: bZn conn den}
\end{aligned}
\end{equation}
Finally, using~\eqref{eq: bZn conn num} and~\eqref{eq: bZn conn den}, we find that the connected part of the $n$-th moment of the overlaps reads 
\be
\left.\overline{\langle \Psi^E_{k_1} | \Psi^E_{k_2} \rangle \langle \Psi^E_{k_2} | \Psi^E_{k_3} \rangle \ldots \langle \Psi^E_{k_{n}} | \Psi^E_{k_1} \rangle} \right|_{\text{conn.}} = e^{(1-n)(\bS_L+\bS_R)}\,. \label{eq: microoverlap-n}
\ee

The semiclassical approximation of the connected part of the moments of the overlaps captures the amount of non-orthogonality between the semiclassical microstates in~\eqref{eq:Kspanmicro}. In the next section, we will leverage this information to compute the correct dimension of the resulting Hilbert space.

\subsection{Resolvent matrix and counting states}
\label{sec: the resolvent}
Equipped with the moments of overlaps $\left(G^n\right)_{ij}$ computed in \autoref{sec: microoverlaps}, we proceed to compute the dimension of the Hilbert space spanned by the semiclassical microstates in~\eqref{eq:Kspanmicro}. To this end, we introduce the resolvent matrix~\eqref{eq: resolvent}. The trace of the resolvent matrix exhibits poles at each eigenvalue of $G$, with the residue of each pole indicating the degeneracy of the corresponding eigenvalue.  In the semiclassical approximation, we can calculate its trace, which is given by
\begin{equation}\label{eq:resolve-bar}
    \overline{R(\lambda)} = \frac{\Omega}{\lambda} + \frac{1}{\lambda} \sum_{n=1}^\infty  \sum_{i=1}^\Omega \frac{\overline{\left(G^n\right)_{ii}}}{\lambda^n}\,,
\end{equation}
where $\overline{R(\lambda)} = \sum_{i=1}^\Omega \overline{R_{ii}(\lambda)}$, $\Omega$ is the dimension of the Gram matrix, and the moments of the overlaps $\overline{\left(G^n\right)_{ii}}$ can be obtained from eq.~\eqref{eq: bZn}. 

The expansion in~\eqref{eq:resolve-bar} can be represented diagrammatically as
\be
    \includegraphics[scale=0.9]{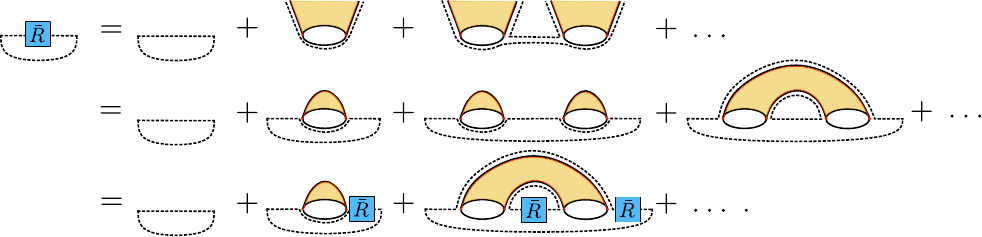} 
    \label{fig:SDR}
\ee
The first line in this diagrammatic expansion depicts each term in the infinite sum over $n$ in~\eqref{eq:resolve-bar}. Each term features $n$ copies of Euclidean time contours (schematically shown by circles) connected by dashed lines representing the microstate indices. Red dots show the operator insertions.  To evaluate the trace of the resolvent via the gravitational path integral, one sums over all geometries consistent with the boundary conditions set by the Euclidean time contours, as shown in the second line. 

We focus on the regime where both $\Omega$ and $e^{1/G_N}$ are large, allowing us to restrict our attention to planar geometries, as explained in~\cite{Penington19}. The sum over all planar geometries can be reorganized to include only contributions from fully connected planar geometries, at the cost of an $\overline{R}$-insertion for each copy of the overlap. This is illustrated in the last line. In this limit, the Schwinger-Dyson equation becomes
\begin{equation}\label{eq:resolve-bar-connected}
    \overline{R} = \frac{\Omega}{\lambda} + \frac{1}{\lambda} \sum_{n=1}^\infty  \sum_{i=1}^\Omega  \frac{\overline{R}^n}{\Omega^n} \overline{\left(G^n\right)_{ii}}\Big|_{\rm conn.}\,.
\end{equation}

In the heavy shell limit ($\mu\to\infty$), $\overline{\left(G^n\right)_{ii}}\Big|_{\rm conn.}$ is given by~\eqref{eq: microoverlap-n} times a factor of $\Omega^n$ coming from the summation over the indices in the Gram matrix products.  The Schwinger-Dyson equation~\eqref{eq:resolve-bar-connected} then reduces to
\begin{equation}\label{eq:eqnresolvent}
     \lambda \overline{R}= \Omega + \frac{\overline{R} e^{\bS_L+\bS_R}}{e^{\bS_L+\bS_R} -\overline{R}}\,.
\end{equation}
This is a quadratic equation in $\overline{R}$, yielding two solutions
\begin{equation}\label{eq: Rsols}
\overline{R_\pm(\lambda)}=\frac{e^{\bS_L+\bS_R} \left(\lambda -1\right)+\Omega\pm\sqrt{(e^{\bS_L+\bS_R} (\lambda -1)+\Omega )^2-4 e^{\bS_L+\bS_R} \lambda  \Omega } }{2 \lambda }\,.
\end{equation}
To identify the correct solution, we impose two constraints on $\overline{R}$. First, since the Gram matrix has $\Omega$ eigenvalues, all real and non-negative, $\overline{R}$ needs to satisfy
\begin{equation}\label{eq: R constraint}
    \oint \frac{d\lambda}{2\pi i} \overline{R}= \Omega\,,
\end{equation}
where the contour is counter-clockwise and encircles the line $[0,\Lambda]$ and $\Lambda$ is large enough that all branch points and poles of the resolvent are in the region $|\lambda| < \Lambda$. Second, because the trace of the Gram matrix equals $\Omega$, $\overline{R}$ also needs to satisfy 
\begin{equation}\label{second}
    \oint \frac{d\lambda}{2\pi i} \lambda \overline{R}=\Omega,
\end{equation}
for the same contour as in~\eqref{eq: R constraint}. Of the two solutions in~\eqref{eq: Rsols}, only $\overline{R_+(\lambda)}$ satisfies the second constraint in~\eqref{second}, uniquely identifying it as the correct solution. Moreover, as $\lambda\to 0$, $\overline{R_+(\lambda)}$ behaves as
\begin{equation}\label{eq:asymptoticsR}
\overline{R_+}= \frac{\Omega-e^{\bS_L+\bS_R}}{\lambda}\Theta\left(\Omega-e^{\bS_L+\bS_R}\right)+R_0 + O(\lambda)\,,
\end{equation}
where $\Theta$ is the Heaviside step function and 
\begin{equation}\label{R0}
R_0=\frac{e^{\bS_L+\bS_R} \left(\left(e^{\bS_L+\bS_R}+\Omega \right) \Theta \left(\Omega -e^{\bS_L+\bS_R}\right)-\Omega \right)}{e^{\bS_L+\bS_R}-\Omega } \,.
\end{equation}

For $\Omega\leq e^{\bS_L+\bS_R}$, the solution $\overline{R_+}$ is regular at $\lambda=0$, specifically $\overline{R_+}|_{\lambda=0}=R_0$ with $R_0$ a finite number. This implies that the Gram matrix has no zero eigenvalue and thus all states spanning $\mathcal{H}^E_{\rm bulk}(\Omega)$ in~\eqref{eq:Kspanmicro} are linearly independent. For $\Omega\geq e^{\bS_L+\bS_R}$, the residue of the resolvent $\overline{R}_+$ at $\lambda=0$ is non-zero and equals $\text{ Res}_{\lambda=0}\overline{R}=\Omega-e^{S_L+S_R}$. This corresponds to the number of linearly dependent states and the degeneracy of the zero eigenvalue of the Gram matrix $G$ is $\Omega-e^{\bS_L+\bS_R}$. Hence, increasing the value of $\Omega$ past $e^{\bS_L+\bS_R}$ leads only to an increase in the number of null states in~\eqref{eq:Kspanmicro}, while the number of linearly independent states remains unchanged. This critical value of $\Omega=e^{\bS_L+\bS_R}$ at which null states emerge corresponds to the dimension of the microcanonical Hilbert space $\mathcal{H}^E$.

\section{Counting microstates of macroscopic black hole fluctuations}
\label{sec:mainresult}
The main result of \cite{shells1} is that the dimension of the Hilbert space~\eqref{eq:Kspanmicro} spanned by semiclassical microstates with a single very massive shell equals the exponential of the black hole entropy whenever $\Omega > e^{\bS_L+\bS_R}$. Our goal is to see how this result is modified if there is a macroscopic fluctuation in which a second shell, carrying an order one fraction of the total energy, is emitted from the past horizon and then reabsorbed into the future horizon.  We want to count the microstates that are consistent with the presence of this fluctuation.

\subsection{The CFT definition}
\label{CFT def two}
\begin{figure}[t]
    \centering
    \includegraphics[scale=0.5]{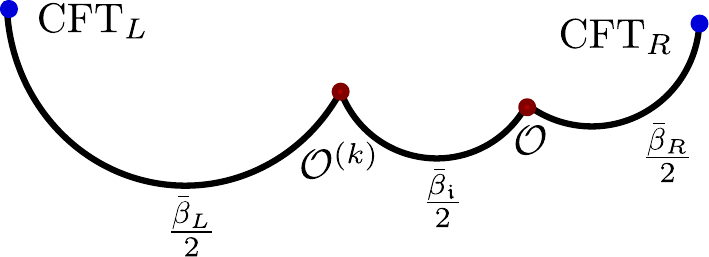} \quad 
    \includegraphics[scale=0.5]{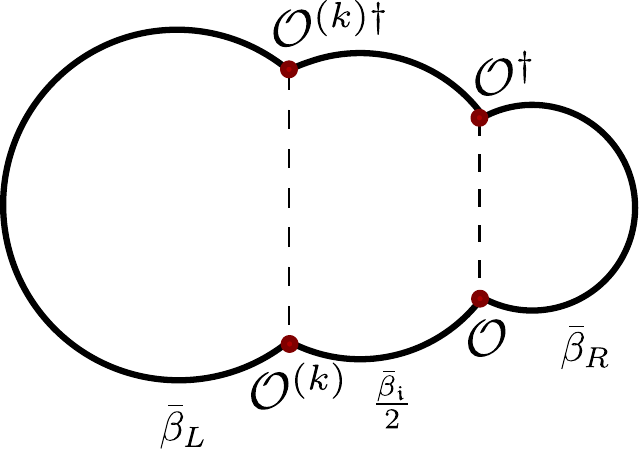} \\ 
    (a) \hspace{5cm} (b)
    \caption{(a) Path integral contour that prepares the state given in eq.~\eqref{state with two shells} The lengths of the circular segments are given by the Euclidean preparation times $\Bar{\beta}_L/2$, $\Bar{\beta}_\i/2$ and $\Bar{\beta}_R/2$. (b) Euclidean time contour which computes the norm squared of the state given in eq.~\eqref{state with two shells}. This quantity can be computed using the gravitational path integral evaluated for geometries with asymptotic boundary length ${\Bar{\beta}_L+\Bar{\beta}_\i + \Bar{\beta}_R}$ and with matter excitations sourced by the operators $\mO$, $\mO^{(k)}$, $\mO^{(k)\dagger}$ and $\mO^\dagger$. In both figures, there is a transverse sphere $\mathbb{S}^{d-1}$ at every point.} 
    \label{fig:contour}
\end{figure}

We can naturally extend the states introduced in~\eqref{eq: single-shell microstates} to include the insertion of two operators. In this section, we focus on the following family of states
\be
| \psi_k \rangle = \frac{1}{\sqrt{\cZ_{kk}}} \sum_{n, m, s} e^{-\frac12 \left(\Bar{\beta}_L E_n + \Bar{\beta}_\i E_s + \Bar{\beta}_R E_m\right)}  \mO^{(k)}_{ms} \mO_{sn}|m \rangle_L \otimes |n \rangle_R\,, \label{state with two shells}
\ee
where the normalization factor is 
\be
\cZ_{kk}=\text{Tr}\left[ \mO^{\dagger} e^{- \frac{1}{2}\Bar{\beta}_\i H } \mO^{(k)\dagger}e^{- \Bar{\beta}_L H }\mO^{(k)} e^{- \frac{1}{2}\Bar{\beta}_\i H }\mO e^{-\Bar{\beta}_R H}\right]\,.\label{eq: Zkk can two shells}
\ee
The operators $\mO^{(k)}$ are the same as those defined in \autoref{sec:review}, with masses specified in~\eqref{eq: mass Ok}. In what follows, these masses are taken to be very large, $\mu\to\infty$,  which is equivalent to considering the limit  $\Bar{\beta}_L\to\beta_L$. The operator $\mO$ is a scalar operator that, in the dual gravitational description, creates a second spherically symmetric thin shell of matter of mass $m$ that will classically backreact on the geometry.  The states can be prepared with Euclidean path integrals performed along the contour shown in \autoref{fig:contour}(a). The norm of the state $\langle \psi_k | \psi_k \rangle$ is computed by a path integral with the Euclidean time contour shown in \autoref{fig:contour}(b).

Similarly to the one shell case, we can project these states onto the microcanonical band $[E_{L,\, R},\, E_{L, \,R} + \delta E)$. We define the projected state to be
\be
\begin{aligned}
|\psi^E_k \rangle & = \frac{1}{\sqrt{\bcZ_{kk}}}  \sum_{|m\rangle_L \otimes |n\rangle_R \in \mH^{E_L}\otimes \mH^{E_R} }  \left(\mO^{(k)} \Pi_{E_\i} \mO\right)_{mn} |m\rangle_L \otimes |n\rangle_R\,,\\
& = \frac{1}{\sqrt{\bcZ_{kk}}}  \sum_{|m\rangle_L \otimes |n\rangle_R \in  \mH^{E_L}\otimes \mH^{E_R} } \sum_{|s\rangle \in \mH^{E_\i}}  \mO^{(k)}_{ms} \mO_{sn} |m\rangle_L \otimes |n\rangle_R\,,\label{eq: microstate two shells}
\end{aligned}
\ee
we additionally projected the intermediate energy into a microcanonical band $[E_\i, E_\i+\delta E)$ using the microcanonical projector $\Pi_{E_\i}$. We introduced the shorthand notation $E=(E_L, E_\i, E_R)$ to indicate that the states $|\psi^E_k\rangle$ are defined by some fixed energies $E_L, E_\i$ and $E_R$. As usual, the states are normalized by\footnote{We include the energy labels here to specify that we are working at fixed energies $E_L, E_\i$ and $E_R$. For clarity, we will omit the energy labels for the rest of the section.} 
\begin{align}\label{eq: bZkk}
{\bcZ_{kk}} (E_L,E_\i,E_R) &=\text{Tr}\left[\Pi_{E_L}\mO^{(k)} \Pi_{E_\i}\mO\Pi_{E_R} \mO^{\dagger} \Pi_{E_\i} \mO^{(k)\dagger} \right] \,,
\end{align}
which can be computed from the canonical normalization factor $\cZ_{kk}$ by an inverse Laplace transform as 
\begin{equation}
    \bcZ_{kk} = \frac{\bch_{kk}}{4\pi^2 } \int  d\bar{\beta}_L d\bar{\beta}_\i d\bar{\beta}'_\i d\bar{\beta}_R \, e^{\bar{\beta}_L E_L +\frac{1}{2}\left(\bar{\beta}_\i + \bar{\beta}'_\i \right)E_\i + \bar{\beta}_R E_R} \cZ_{kk}\left(\bar{\beta}_L,\bar{\beta}_\i,\bar{\beta}'_\i,\bar{\beta}_R\right)\,,\label{eq: bcZkk}
\end{equation}
where $\cZ_{kk}\left(\bar{\beta}_L,\bar{\beta}_\i,\bar{\beta}'_\i,\bar{\beta}_R\right)$ is a generalization of~\eqref{eq: Zkk can two shells} where $\bar{\beta}_\i$ and $\bar{\beta}'_\i$ are taken to be independent and $\bch_{kk}$ is the Hessian determinant of $-\log  \cZ_{kk}\left(\bar{\beta}_L,\bar{\beta}_R,\bar{\beta}_\i,\bar{\beta}'_\i\right)$ with respect to $\Bar{\beta}_{L,R}$, $\bar{\beta}_\i$ and $\bar{\beta}'_\i$ evaluated at the saddle point value.

In what follows, we aim to compute the dimension of the space spanned by $\Omega$ of the microcanonical states $| \psi^E_k \rangle$ 
\begin{equation}\label{eq:Kspanmicro2}
    {\cal H}^E_{\rm bulk}(\Omega) \equiv {\rm Span}\{|\psi^E_k\rangle, \quad k=1,\ldots,\Omega\}
\end{equation}
as a function of the mass $m$ of the additional shell. Notice that from a boundary point of view, one can argue that the dimension of ${\cal H}^E_{\rm bulk}(\Omega)$ is upper bounded by $\text{min}\left(\rm{dim}\left(\mathcal{H}_L\right)\times \rm{dim}\left(\mathcal{H}_R\right), \rm{dim}\left(\mathcal{H}_L\right)\times \rm{dim}\left(\mathcal{H}_\i\right)\right)$. These bounds follow from counting the number of independent coefficients in the definition of the states~\eqref{eq: microstate two shells} that are associated with the operator $\mO^{(k)}$. Specifically, the first bound arises because the matrix $ \left(\mO^{(k)} \Pi_{E_\i} \mO\right)_{mn}$ can lead to maximally $\rm{dim}\left(\mathcal{H}_L\right)\times \rm{dim}\left(\mathcal{H}_R\right)$ independent coefficients since the index $m$ is restricted to run over  $\rm{dim}\left(\mathcal{H}_L\right)$ different states and similarly the index $n$ runs over $\rm{dim}\left(\mathcal{H}_R\right)$ different states. The second bound is found analogously and arises because the matrix $\mO^{(k)}_{ms}$ can give to at most $\rm{dim}\left(\mathcal{H}_L\right)\times \rm{dim}\left(\mathcal{H}_\i\right)$ independent coefficients because the index $s$ runs over $\rm{dim}\left(\mathcal{H}_\i\right)$ different states. Assuming that the matrices associated with the operator $\mO^{(k)}$ are random, consistent with an ETH ansatz, will ensure that the dimension of the Hilbert space saturates this bound. 

We will verify this bound and relate the dimension to the black hole entropy using a gravity calculation. Specifically, we will carefully compute the dimension of the Hilbert space spanned by such states as a function of the mass of the additional shell $m$. First, we consider states for which the additional shell is very massive, $m>m_0$, where $m_0$ is defined in eq.~\eqref{eq: m_0}. Second, we explore the dimension of the Hilbert space spanned by shells with small mass $m \ll m_0$. For concreteness, we focus on the case of $d=2$ and comment on how the results extend to general dimensions in the discussion.

\subsection{Large mass regime: a second internal shell}\label{sec: large mass}

We begin with the regime in which the additional shell has a large mass, $m>m_0$, so that the dominant contributions to the saddle point approximation of the inverse Laplace transform are geometries in which both shells are inside the horizons of the two-sided black hole. 
The connected contribution to the moments of overlaps between the states in the family~\eqref{eq:Kspanmicro2} can be computed semiclassically as outlined in~\autoref{sec: microoverlaps} and is given by 
\begin{equation}\label{eq: nth overlap two}
    \left. \overline{\langle \psi^E_{k_1} | \psi^E_{k_2} \rangle \langle \psi^E_{k_2} | \psi^E_{k_3} \rangle \ldots \langle \psi^E_{k_{n}} | \psi^E_{k_1} \rangle}\right|_{\rm conn.} = \frac{ \left.\overline{\bcZ_{k_1 k_2}\bcZ_{k_2 k_3}\ldots \bcZ_{k_{n} k_1}}\right|_{\rm conn.}}{\overline{\bcZ_{k_1 k_1}\bcZ_{k_2 k_2} \ldots \bcZ_{k_{n} k_{n}}}}\,,
\end{equation}
where 
\begin{equation}
\begin{aligned}
\overline{\bcZ_{i_1j_1} \ldots \bcZ_{i_nj_n}}=& \frac{\mathbf{\mathfrak{H}}}{\left(4\pi^2 \right)^n} \int  \overline{\cZ^{(1)}_{i_1j_1} \ldots \cZ^{(n)}_{i_nj_n}}\times\\
&\qquad\qquad\prod_{i}^n \left( d\bar{\beta}^{(i)}_L d\bar{\beta}^{(i)}_\i d\bar{\beta}^{'(i)}_\i d\bar{\beta}^{(i)}_R \,  e^{\bar{\beta}^{(i)}_L E_L +\frac{n}{2}\left(\bar{\beta}^{(i)}_\i + \bar{\beta}^{'(i)}_\i \right)E_\i + \bar{\beta}^{(i)}_R E_R}\right)\,,
\end{aligned}  \label{eq: bcZij}
\end{equation}
where $\cZ^{(i)}_{ij} = \cZ_{ij}\left(\bar{\beta}_L^{(i)},\bar{\beta}_R^{(i)},\bar{\beta}_\i^{(i)},\bar{\beta}_\i^{'(i)}\right)$ and $\mathbf{\mathfrak{H}}$ is the Hessian determinant of $-\log  \overline{\cZ^{(1)}_{i_1j_1} \ldots \cZ^{(n)}_{i_nj_n}} $ with respect to $\Bar{\beta}^{(i)}_{L,R}$, $\bar{\beta}^{(i)}_\i$ and $\bar{\beta}^{'(i)}_\i$ evaluated at the saddle point value.

\paragraph{First moment.}
\begin{figure}[t]
    \centering
        \includegraphics[scale=0.70]{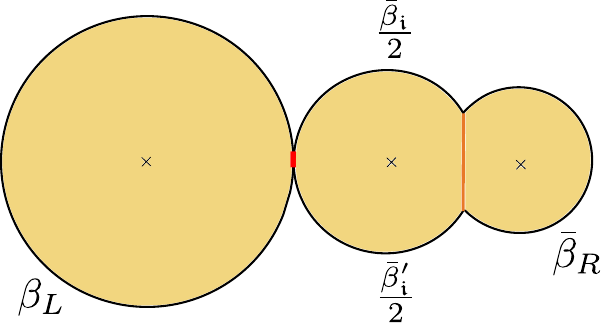} 
    \caption{The geometry computing the overlaps $\overline{\langle \psi_k | \psi_k \rangle}$. The vertical red and orange lines represent the two shell worldvolumes. We consider the limit where the orange shell has $m> m_0$ such that it is behind the black hole horizon $r_R$. The red shell has large mass $\mu\to \infty$ which ensures that the red shell worldvolume is behind $r_L$ and pinches off, $\Bar{\beta}_{L}\to \beta_{L}$.} 
    \label{fig:2internalshell2}
\end{figure}

Let us consider the overlap $\overline{\langle \psi_i^E | \psi_j^E \rangle}$. To this end, we need to compute $\overline{\bcZ_{ij}}$ using $\overline{\cZ_{ij}\left(\bar{\beta}_L,\bar{\beta}_\i,\bar{\beta}'_\i,\bar{\beta}_R\right)}$ calculated using a semiclassical approximation to the gravity path integral. In turn, the latter computation amounts to finding the classically allowed geometries that fill in the boundary conditions specified in~\autoref{fig:contour}(b) and computing their on-shell action. As before, if $i \neq j$, no solution of the gravity equations of motion satisfies the boundary conditions given by the contour in~\autoref{fig:contour}(b). Furthermore, since the states are normalized, we find $\overline{\langle \psi^E_i | \psi^E_j \rangle}  = \delta_{ij}$.

As an instructive exercise, we compute the on-shell actions that normalize the states in~\eqref{eq: microstate two shells}. When the masses of the operators $\mO^{(i)\dagger}$ and $\mO^{(j)}$ are equal, $m_i=m_j=i\mu$, where $\mu$ is taken to be large, both shells are inside both horizons, and the geometry is given in~\autoref{fig:2internalshell2}. The inverse temperatures $\beta_L$, $\beta_\i$ and $\beta_R$ in the different sections of the geometry are given by
\begin{equation}
    \Bar{\beta}_L=\beta_L\,, \quad \frac{\Bar{\beta}_\i+\Bar{\beta}_\i'}{2} = \beta_\i-\frac{\beta_\i}{\pi}\arcsin\left(\frac{r_\i}{R_*}\right)
    \,, \quad \Bar{\beta}_R=\beta_R-\frac{\beta_R}{\pi}\arcsin\left(\frac{r_R}{R_*}\right)\,,
\end{equation}
where the turning point of the shell with mass $m$ is given by 
\begin{equation}
    R_*^{2}=r_R^{2}+\left(\frac{M_R-M_\i}{m}-2G m\right)^2\,.    
\end{equation}

The on-shell action that computes $\overline{\cZ_{kk}\left(\bar{\beta}_L,\bar{\beta}_\i,\bar{\beta}'_\i,\bar{\beta}_R\right)}$ is given by
\be
I_{\rm tot}^{\rm ren}\Big|_{\rm on-shell}=\IPL{\,L}^{\rm ren}+\ICL{\,L}^{\rm ren}+I_{\textcolor{red}{\cal W}}^{\rm ren} +\ICR{\,\i}^{\rm ren}+\IPiR{\,\i}^{\rm ren}+\IPiR{\i'}^{\rm ren}+\ICL{\,\i}^{\rm ren}+I_{\textcolor{orange}{\cal W}}^{\rm ren} +\ICR{\,R}^{\rm ren} +\IPR{\,R}^{\rm ren}\,,
\ee
where the various contributions are given in \autoref{sec: building blocks}. Specifically, we find that in the pinching limit $\mu\to\infty$, the saddle point approximation to the normalization coefficient gives
\begin{equation}
\begin{aligned}
    -\log  \overline{\cZ_{kk}} =& \Bar{\beta}_L \mathcal{F}(\Bar{\beta}_L) + 2\mu \log \frac{1}{G \mu} +\frac{\bar{\beta}_\i+\bar{\beta}_\i'}{2} \mathcal{F}(\beta_\i) \\
    &+ \frac{\sqrt{R_*^2-r_R^2}+\sqrt{R_*^2-r_\i^2}}{2G} \log\frac{2}{R_*}+\bar{\beta}_R \mathcal{F}(\beta_R)\,.
\end{aligned}
\end{equation}
Recall that $\mathcal{F}(\beta)$ is the free energy defined in~\eqref{eq:renormF}. In the large $m$ regime, the relation between the Euclidean lengths $\bar{\beta}_\i$, $\bar{\beta}'_\i$, $\bar{\beta}_R$ and the black hole temperatures $\beta_\i$ and $\beta_R$ is
\begin{equation}
    \beta_\i = \frac{\bar{\beta}_\i +\bar{\beta}'_\i}{2} +\frac{1}{G m} -\frac{ \overline{M}_i+3 \overline{M}_R}{6G^2   m^3}+\mathcal{O}\left(m^{-4}\right)
   \,, \quad\beta_R = \bar{\beta}_R +\frac{1}{G m} -\frac{ \overline{M}_R+3 \overline{M}_\i}{6G^2   m^3}+\mathcal{O}\left(m^{-4}\right)\,,
\end{equation}
where $\overline{M}_\i = \frac{2\pi^2}{G (\bar{\beta}_\i+\bar{\beta}'_\i)^2}$ and $\overline{M}_R = \frac{\pi^2}{2G \bar{\beta}_R^2}$. The on-shell action of the geometry computing the norm of the state in this regime is therefore
\begin{equation}\label{eq: onshell action two shells large mass norm}
\begin{aligned}
    -\log  \overline{\cZ_{kk}} = &\bar{\beta}_L \mathcal{F}(\bar{\beta}_L) +\frac{\bar{\beta}_\i +\bar{\beta}'_\i}{2}  \mathcal{F}\left(\frac{\bar{\beta}_\i +\bar{\beta}'_\i}{2} \right) + \bar{\beta}_R \mathcal{F}(\bar{\beta}_R) +2 {m} \log \frac{1}{G {m}} +2 \mu \log \frac{1}{G \mu}\,\\
    &+ \frac{1}{G} \frac{\overline{M}_\i+\overline{M}_R}{m}- \frac{1}{G^2m^2} \left( \frac{2\overline{M}_\i}{\bar{\beta}_\i+\bar{\beta}'_\i}+\frac{\overline{M}_R}{\bar{\beta}_R}\right)+\mathcal{O}\left(m^{-3}\right) \,.
\end{aligned}
\end{equation}

The saddle point values of the Euclidean lengths are given by
\begin{equation}
    \bar{\beta}_L = \frac{\pi^2}{G  \bS_L} \,,\quad \frac{\bar{\beta}_\i+\bar{\beta}_\i'}{2} = \frac{\pi^2}{G  \bS_\i} -\frac{1}{ G m}+\mathcal{O}\left(m^{-3}\right)\,,
    \quad \bar{\beta}_R = \frac{\pi^2}{G  \bS_R} -\frac{1}{ G m}+\mathcal{O}\left(m^{-3}\right)\,,
\end{equation}
where 
\begin{equation}\label{eq: entropy}
    \bS_{\alpha} = \pi \sqrt{\frac{2 E_{\alpha} }{G}}\,, \quad \alpha = L,\i,R
\end{equation}
is the entropy of a black hole with energy $E_{\alpha}$. Therefore, the saddle point approximation of the normalization constant is
\begin{equation}\label{eq:bZkk}
\overline{\bcZ_{kk} } = e^{-\bI_1}\,,
\end{equation}
to leading order in the semiclassical approximation, where
\begin{equation}
    \bI_1 = - \bS_L - \bS_\i -\bS_R +2 {m} \log \frac{1}{G {m}} +2 \mu \log \frac{1}{G \mu}+  \frac{ \left(\bS_R^2+\bS_\i^2\right)}{2\pi ^2 m}+\mathcal{O}\left(m^{-3}\right) \,.
\end{equation} 

\paragraph{Higher moments.} 

\begin{figure}[t]
    \centering
    \includegraphics[scale=0.4]{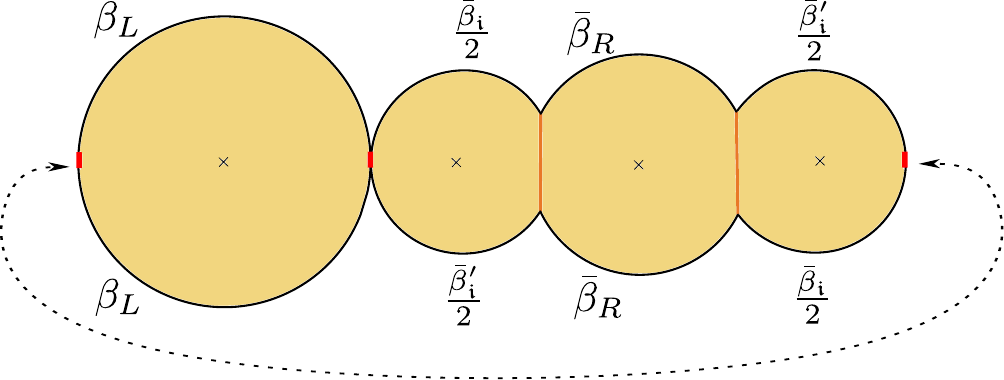} \quad 
    \includegraphics[scale=0.4]{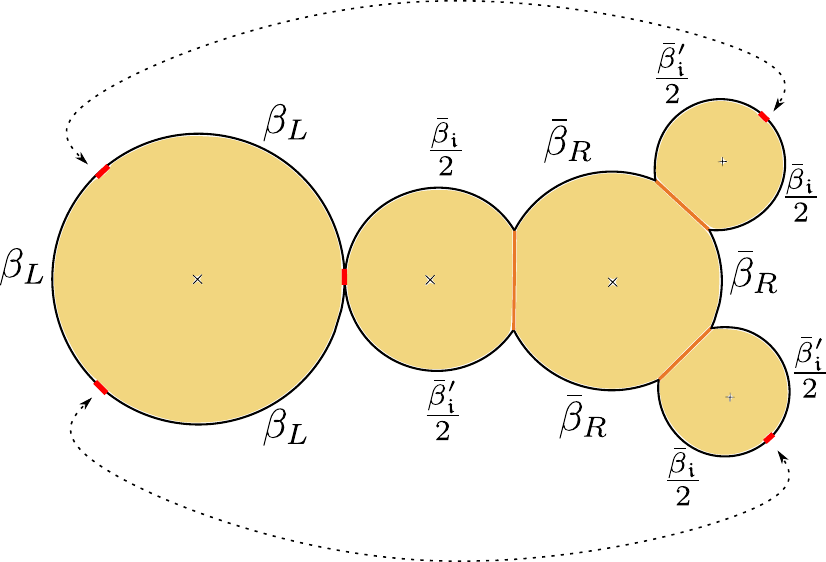} \\ 
    (a) \hspace{5cm} (b)
    \caption{Wormhole geometries which dominate the connected contributions to moments of overlaps of the states in~\eqref{eq: microstate two shells} with large mass $m$ when $E_\i$ is greater than $E_{R}$. (a) Two-boundary wormhole which computes the connected contribution to $\overline{\cZ_{ij}\cZ_{ji}}$. (b) Three-boundary wormhole which computes the connected contribution to $\overline{\cZ_{ik}\cZ_{kj}\cZ_{ji}}$. Both geometries are given in the pinching limit $\mu\to\infty$ for which $\Bar{\beta}_L\to\beta_L$.}  
    \label{fig:caterpillar}
\end{figure}

\begin{figure}[t]
    \centering
    \includegraphics[scale=0.65]{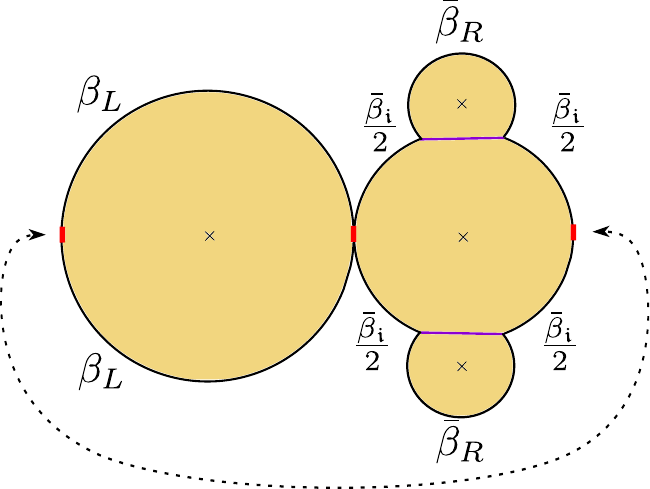} \quad 
    \includegraphics[scale=0.65]{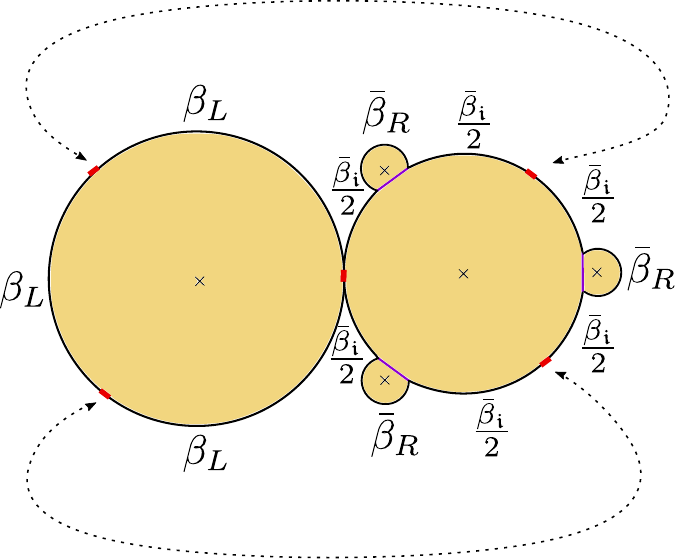} \\ 
    (a) \hspace{5cm} (b)
    \caption{Wormhole geometries which dominate the connected contributions to moments of overlaps of the states in~\eqref{eq: microstate two shells} with large mass $m$ in the regime where $E_R$ is greater than $E_\i$. (a) Two-boundary wormhole which computes the connected contribution to $\overline{\cZ_{ij}\cZ_{ji}}$. (b) Three-boundary wormhole which computes the connected contribution to $\overline{\cZ_{ik}\cZ_{kj}\cZ_{ji}}$. Both geometries are given in the pinching limit $\mu\to\infty$ for which $\Bar{\beta}_L\to\beta_L$.} 
    \label{fig:butterflyref}
\end{figure}

The higher moments of overlaps can be computed analogously to the procedure outlined in~\autoref{sec: microoverlaps}. To this end, we need to compute products of $\cZ_{i j}$ as discussed in~\autoref{sec: Overlaps}. As usual, it is sufficient to consider only the connected contributions for later purposes and compute $ \left.\overline{\cZ_{k_1 k_2}\cZ_{k_2 k_3}\ldots \cZ_{k_{n} k_1}}\right|_{\rm conn.}$. The dominant contributions depend on the relation between the energies $E_R$ and $E_\i$. The geometries for $n=2$ and $n=3$ in the limit where $\mu\to\infty$ which are relevant for the case where $E_\i$ is greater than $E_{R}$ are shown in~\autoref{fig:caterpillar}, while those for which $E_\i$ is smaller than $E_R$ are in~\autoref{fig:butterflyref}. 

First, consider $E_\i$ to be greater than $E_R$. In this regime, the temperatures of the various regions are related to the Euclidean preparation times by
\begin{equation}
    n q_L=\Tilde{\beta}_L\,,\quad \frac{\Bar{\beta}_\i^{(i)}+\Bar{\beta}_\i'^{(i+1)}}{2}= \Tilde{\beta}_\i^{(i)}-\frac{\Tilde{\beta}_\i^{(i)}}{\pi}\arcsin\left(\frac{\Tilde{r}_\i^{(i)}}{\Tilde{R}_*^{(i)}}\right)\,,\quad n q_R=\Tilde{\beta}_R-\sum_{i=1}^n \frac{\Tilde{\beta}_R}{\pi}\arcsin\left(\frac{\Tilde{r}_R}{\Tilde{R}_*^{(i)}}\right)\,,
    \end{equation}
where we have used $q_{L,R} = \frac{1}{n}\sum_i \Bar{\beta}^{i}_{L,R}$ and we have identified $\Bar{\beta}_\i'^{(n+1)}=\Bar{\beta}_\i'^{(1)}$. The turning points of the shells of mass $m$ are
\begin{equation}
\Tilde{R}_*^{(i)2}= \Tilde{r}_R^2+\left(\frac{\Tilde{M}_R-\Tilde{M}_\i^{(i)}}{m}-2G m\right)^2\,,
\end{equation}
where $\tilde{M}_{\i}^{(i)} = \frac{\pi^2}{2G \Tilde{\beta}_{\i}^{(i)2}}$ and $\tilde{M}_{R} = \frac{\pi^2}{2G \Tilde{\beta}_{R}^2}$. The corresponding on-shell actions lead to 
\begin{equation}
\begin{aligned}
    - \log  \left.\overline{\cZ_{k_1 k_2}\ldots \cZ_{k_{n} k_1}}\right|_{\rm conn.} = & n q_L \mathcal{F}(\tilde{\beta}_L) +\sum_{i=1}^n \frac{\bar{\beta}_\i^{(i)}+\bar{\beta}_\i'^{(i+1)}}{2} \mathcal{F}(\tilde{\beta}_\i^{(i)}) + n q_R \mathcal{F}(\tilde{\beta}_R) \\
    & + \sum_{i=1}^n\frac{\sqrt{R_*^{(i)2}-\tilde{r}_R^2}+\sqrt{R_*^{(i)2}-\tilde{r}_\i^{(i)2}}}{2G} \log\frac{2}{R_*^{(i)}}+2 \mu n \log \frac{1}{G \mu} \,.
\end{aligned}
\end{equation}

In the large mass $m$ regime,
the Euclidean lengths $\bar{\beta}^{(i)}_\i$, $\bar{\beta}^{'(i)}_\i$ and the average Euclidean lengths $q_L$ and $q_R$ are related to the temperatures $\tilde{\beta}_L$, $\tilde{\beta}^{(i)}_\i$ and $\tilde{\beta}_R$ via
\begin{equation}
\begin{aligned}
    \Tilde{\beta}_L&=nq_L\,,\\\tilde{\beta}_\i^{(i)} &= \frac{\bar{\beta}_\i^{(i)}+\bar{\beta}_\i'^{(i+1)}}{2} +\frac{1}{G m} -\frac{ n^2\overline{M}_\i^{(i)}+3 \overline{M}_R}{6G^2n^2   m^3}+\mathcal{O}\left(m^{-4}\right)\,,\\ 
    \frac{\tilde{\beta}_R}{n} & = q_R +\frac{1}{G m} - \frac{1}{n} \sum_{i=1}^n\frac{ \overline{M}_R+3 n^2\overline{M}_\i^{(i)}}{6G^2 n^2  m^3}+\mathcal{O}\left(m^{-4}\right),
\end{aligned}
\end{equation}
where $\overline{M}_\i^{(i)} = \frac{2\pi^2}{G (\bar{\beta}_\i^{(i)}+\bar{\beta}_\i'^{(i+1)})^2}$ and $\overline{M}_R = \frac{\pi^2}{2G q_R^2}$. In the large $m$ regime, the on-shell action of the geometry computing $\left.\overline{\cZ_{k_1 k_2}\ldots \cZ_{k_{n} k_1}}\right|_{\rm conn.} $ is
\begin{equation}
\begin{aligned}
    - \log  \left.\overline{\cZ_{k_1 k_2}\ldots \cZ_{k_{n} k_1}}\right|_{\rm conn.} =&n q_L \mathcal{F}(n q_L) +\sum_i^n\frac{\bar{\beta}_\i^{(i)}+\bar{\beta}_\i'^{(i+1)}}{2} \mathcal{F}\left(\frac{\bar{\beta}_\i^{(i)}+\bar{\beta}_\i'^{(i+1)}}{2}\right) + n q_R \mathcal{F}(n q_R) 
    \\ & +2n {m} \log \frac{1}{G {m}} + 2n\mu \log \frac{1}{G \mu}+ \frac{1}{G} \sum_{i=1}^n \frac{\overline{M}_R+\overline{M}_\i^{(i)} n^2}{n^2 m}\\
    & -\frac{1}{G^2m^2} \sum_{i=1}^n \left(\frac{\overline{M}_R}{n^2q_R} + \frac{2\overline{M}_\i^{(i)}}{\bar{\beta}_\i^{(i)}+\bar{\beta}_\i'^{(i+1)}}\right)+\mathcal{O}\left(m^{-3}\right)\,.
\end{aligned}
\end{equation}

The saddle point values of the Euclidean lengths are given by
\begin{equation}
    q_L = \frac{\pi^2}{G n \bS_L} \,,\quad \frac{\bar{\beta}_\i^{(i)}+\bar{\beta}_\i'^{(i+1)}}{2} = \frac{\pi^2}{G  \bS_\i} -\frac{1}{ G m}+\mathcal{O}\left(m^{-3}\right)\,,
    \quad q_R = \frac{\pi^2}{G n \bS_R} -\frac{1}{ G m}+\mathcal{O}\left(m^{-3}\right)\,,
\end{equation}
where we recall that $\bS_{L}$, $ \bS_\i$ and $\bS_R$ are the entropies of black holes with energies $E_L$, $E_\i$ and $E_R$, respectively, as defined in~\eqref{eq: entropy}. Therefore, the saddle point approximation to the numerator of~\eqref{eq: nth overlap two} is
\begin{equation}
\left.\overline{\bcZ_{k_1 k_2}\ldots \bcZ_{k_{n} k_1}}\right|_{\rm conn.} = e^{-\bI_n}\,,
\end{equation}
to leading order in the semiclassical approximation, where in the large $m$ regime one finds
\begin{equation}
    \bI_n= - \bS_L -n \bS_\i -\bS_R +2n {m} \log \frac{1}{G {m}} +2n \mu \log \frac{1}{G \mu}+  \frac{ n\left(\bS_R^2+\bS_\i^2\right)}{2\pi ^2 m}+\mathcal{O}\left(m^{-3}\right) \,.
\end{equation}
The denominator of~\eqref{eq: nth overlap two} is dominated by the fully disconnected contribution, and the corresponding semiclassical approximation is
\begin{equation}\label{eq: bZ denom}
    \overline{\bcZ_{k_1 k_1}\bcZ_{k_2 k_2} \ldots \bcZ_{k_{n} k_{n}}} = e^{-n \bI_1}\,.
\end{equation}
 With these two results, we find the semiclassical approximation to the connected component of the $n$-moment of overlaps is\footnote{While we explicitly display the derivation up to ${\cal O}(m^{-3})$, we have verified it up to ${\cal O}(m^{-8})$. We expect the result to be exact as long as the mass of the shell is large enough ($m>m_0$) that geometries with the shell inside the horizon dominate the inverse Laplace transform.}
\be
\left.\overline{\langle \psi^E_{k_1} | \psi^E_{k_2} \rangle \langle \psi^E_{k_2} | \psi^E_{k_3} \rangle \ldots \langle \psi^E_{k_{n}} | \psi^E_{k_1} \rangle} \right|_{\text{conn.}} = e^{(1-n)(\bS_L+\bS_R)+ {\cal O}(m^{-3})}\,, \quad {\rm for} \ E_\i>E_R\,. \label{eq: microoverlap-n two}
\ee

Next, we turn our attention to the case when $E_\i<E_R$. In this regime, the on-shell action of the geometry computing $\left.\overline{\cZ_{k_1 k_2}\ldots \cZ_{k_{n} k_1}}\right|_{\rm conn.} $  is
\begin{equation}
\begin{aligned}
    - \log  \left.\overline{\cZ_{k_1 k_2}\ldots \cZ_{k_{n} k_1}}\right|_{\rm conn.} = & n q_L \mathcal{F}(\tilde{\beta}_L) +n \frac{q_\i + q_\i'}{2} \mathcal{F}(\tilde{\beta}_\i) + \sum_i^n \bar{\beta}^{(i)}_R \mathcal{F}(\tilde{\beta}^{(i)}_R) \\
    & + \sum_i^n\frac{\sqrt{R_*^{(i)2}-\tilde{r}_\i^2}+\sqrt{R_*^{(i)2}-\tilde{r}_R^{(i)2}}}{2G} \log\frac{2}{R_*^{(i)}}+2 \mu n \log \frac{1}{G \mu} \,,
\end{aligned}
\end{equation}
where in addition to using $q_{L} = \frac{1}{n}\sum_i \Bar{\beta}^{(i)}_{L}$, we have introduced $q_{\i} = \frac{1}{n}\sum_i \Bar{\beta}^{(i)}_{\i}$ and $q'_{\i} = \frac{1}{n}\sum_i \Bar{\beta}^{'(i)}_{\i}$. The temperatures of the various regions are related to the Euclidean preparation times by
\begin{equation}
    n q_L=\Tilde{\beta}_L\,,\quad n\frac{q_\i+q'_\i}{2}= \Tilde{\beta}_\i-\sum_i\frac{\Tilde{\beta}_\i}{\pi}\arcsin\left(\frac{\Tilde{r}_\i}{\Tilde{R}_*^{(i)}}\right)\,,\quad \Bar{\beta}_R^{(i)}=\Tilde{\beta}_R^{(i)}-\frac{\Tilde{\beta}_R^{(i)}}{\pi}\arcsin\left(\frac{\Tilde{r}^{(i)}_R}{\Tilde{R}_*^{(i)}}\right)\,,
    \end{equation}
and the location of the turning point is given by
\begin{equation}
\Tilde{R}_*^{(i)2}=\Tilde{r}_R^{(i)2}+\left(\frac{\Tilde{M}^{(i)}_R-\Tilde{M}_\i}{m}-2G m\right)^2\,.
\end{equation}

The saddle point values of the Euclidean lengths in the limit of large mass $m$ are given by 
\begin{equation}
    q_L = \frac{\pi^2}{G n \bS_L} \,,\quad \frac{q_\i + q'_\i}{2} = \frac{\pi^2}{G n \bS_\i} -\frac{1}{ G m}+\mathcal{O}\left(m^{-3}\right)\,,
    \quad \bar{\beta}_R^{(i)}  = \frac{\pi^2}{G \bS_R} -\frac{1}{ G m}+\mathcal{O}\left(m^{-3}\right)\,,
\end{equation}
where we recall that $\bS_{L}$ is the entropy of a black hole with energy $E_{L}$ as defined in \eqref{eq: entropy}, and similarly $\bS_\i$ and $\bS_R$ are the entropies of black holes with energies $E_\i$ and $E_R$, respectively. The saddle point approximation to the term in the numerator of~\eqref{eq: nth overlap two} is therefore
\begin{equation}
\left.\overline{\bcZ_{k_1 k_2}\ldots \bcZ_{k_{n} k_1}}\right|_{\rm conn.} = e^{-\bI_n}\,,
\end{equation}
to leading order in the semiclassical approximation, where in the large $m$ regime one finds
\begin{equation}
    \bI_n= - \bS_L - \bS_\i - n \bS_R +2n {m} \log \frac{1}{G {m}} +2n \mu \log \frac{1}{G \mu}+  \frac{ n\left(\bS_R^2+\bS_\i^2\right)}{2\pi ^2 m}+\mathcal{O}\left(m^{-3}\right) \,.
\end{equation}
The term in the denominator of~\eqref{eq: nth overlap two} is dominated by the fully disconnected contribution, and the corresponding semiclassical approximation is given by~\eqref{eq: bZ denom}. The resulting semiclassical approximation to the connected component of the $n$-moment of overlaps is
\be
\left.\overline{\langle \psi^E_{k_1} | \psi^E_{k_2} \rangle \langle \psi^E_{k_2} | \psi^E_{k_3} \rangle \ldots \langle \psi^E_{k_{n}} | \psi^E_{k_1} \rangle} \right|_{\text{conn.}} = e^{(1-n)(\bS_L+\bS_\i)+ {\cal O}(m^{-3})}\,, \quad {\rm for} \ E_R>E_\i \label{eq: microoverlap-n two ref}\,,
\ee
which is similar to~\eqref{eq: microoverlap-n two} but with the smaller $\bS_\i$ replacing $\bS_R$.

\paragraph{Hilbert space dimension.} 
For $E_\i>E_R$, we found that the connected contribution to the moments of overlaps~\eqref{eq: microoverlap-n two} is the same as in the single shell states~\eqref{eq: microoverlap-n} up to $\mathcal{O}\left(m^{-3}\right)$. This implies that the Gram matrix of the states in~\eqref{eq:Kspanmicro2} is equal to the one in~\autoref{sec:review} up to possible corrections of ${\cal O}\left(m^{-3}\right)$. On the other hand, for $E_\i<E_R$, the connected contribution to the moments of overlaps~\eqref{eq: microoverlap-n two ref} is similar with $\bS_R$ replaced with the smaller $\bS_\i$. The two results can be summarized by 
\be
\left.\overline{\langle \psi^E_{k_1} | \psi^E_{k_2} \rangle \langle \psi^E_{k_2} | \psi^E_{k_3} \rangle \ldots \langle \psi^E_{k_{n}} | \psi^E_{k_1} \rangle} \right|_{\text{conn.}} = e^{(1-n)(\bS_L+{\rm min}\left(\bS_\i,\bS_R\right))+ {\cal O}(m^{-3})}\,. \label{eq: microoverlap-n two min}
\ee
Therefore, the results of \autoref{sec: the resolvent} can be generalized to the Gram matrix of states with a second shell with fixed large mass. In particular, the rank of the matrix is
\begin{equation}\label{eq:rankG large mass}
{\rm Rank} \left(G_{ij}\right) = 
    \begin{cases}
        \Omega\quad \quad \quad \quad \quad \quad \quad\ \quad \quad {\rm for}  \quad \Omega \leq {\rm min} \left(e^{\bS_L+\bS_R},e^{\bS_L+\bS_\i}\right) \,, \\
        {\rm min} \left(e^{\bS_L+\bS_R},e^{\bS_L+\bS_\i}\right) \quad \ {\rm for} \quad \Omega \geq {\rm min} \left(e^{\bS_L+\bS_R},e^{\bS_L+\bS_\i}\right) \,,
    \end{cases}
\end{equation}
up to possible corrections for finite $m$.\footnote{Note that the stability of the resolvent method against corrections is not well understood. Specifically, the rank of a matrix is unstable against tiny corrections if there are zero
eigenvalues, i.e., if the matrix has less than maximal rank. While we have control of various orders in the $1/m$ expansion, one might wonder whether higher order corrections could change the rank of the matrix dramatically, by adding tiny corrections to many previously vanishing values. A similar subtlety was raised in~\cite{Balasubramanian:2024yxk} regarding higher order corrections to the semiclassical approximation. However, if this
approach indeed provides an adequate account of the Bekenstein-Hawking entropy~\cite{shells1,shells2}, we should expect that the effects of such corrections on the rank of the matrix should be small. We leave it to future work to check this. \label{foot: stability}} As argued in~\autoref{sec:review}, the maximal rank of the Gram matrix corresponds to the dimension of the microcanonical Hilbert space, which is therefore ${\rm min} \left(e^{\bS_L+\bS_R},e^{\bS_L+\bS_\i}\right)$. For $E_\i>E_R$, this is in agreement with the expectations from~\cite{shells1}, that adding a second shell with large mass $m$ behind the horizon to the one shell states used in~\autoref{sec:review} does not change the span of the Hilbert space.  

For $E_\i<E_R$, the dimension of the Hilbert space spanned by these states is $e^{\bS_L+\bS_\i}$, which is a lower dimensional subspace of the black hole Hilbert space since $\bS_\i<\bS_R$, see eq.~\eqref{eq: entropy}. This seems surprising since these states resemble a double-sided black hole with energies $E_L$ and $E_R$ as seen by a simple observer on the outside. Nevertheless, this is consistent with the counting argument given below eq.~\eqref{eq:Kspanmicro2}. In what follows, we consider a different basis of states that resemble microcanonical black holes that emit a portion of their energy as some spherical distribution of matter or radiation, which also spans a subspace of the Hilbert space with dimension $e^{\bS_L+\bS_\i}$.

\subsection{Small mass regime: an emitted shell}

\label{sec: small mass}
\begin{figure}[t]
    \centering
    \includegraphics[scale=0.7]{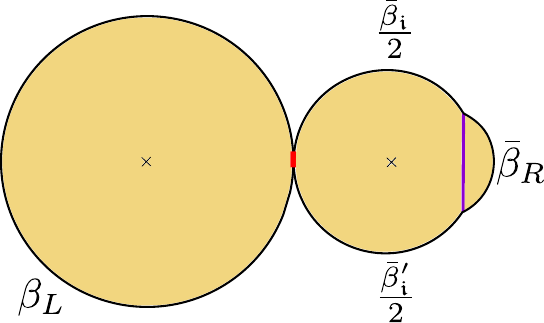} \quad 
    \includegraphics[scale=0.7]{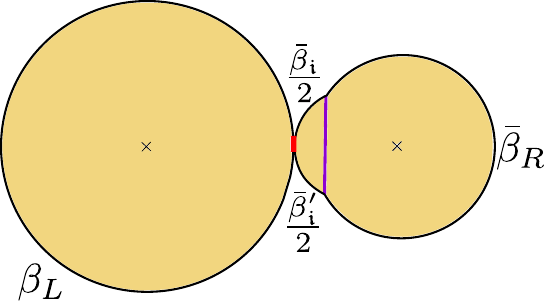} \\ 
    (a) \hspace{5cm} (b)
    \caption{(a) The geometry computing the overlaps $\overline{\langle \psi_k | \psi_k \rangle}$ when $E_R$ is greater than $E_\i$. The vertical red and purple lines represent the two shell worldvolumes. We consider the limit where the purple shell has $m < m_0$ such that it is between the black hole horizon $r_\i$ and the asymptotic boundary. The red shell has large mass $\mu\to \infty$, which ensures that the red shell worldvolume is behind $r_L$ and pinches off and $\Bar{\beta}_{L}\to \beta_{L}$. (b) The geometries that would arise if we take $E_\i>E_{R}$ as well as $\mu\to \infty$ and  $m < m_0$. These geometries lead to similar results to those of~\cite{shells1} and~\autoref{sec: large mass}.} 
    \label{fig:externalshell2}
\end{figure}

We now turn to the regime where the additional shell has a small mass, $m\ll m_0$, and focus on the case where $E_R$ is greater than $E_\i$. For this setup, the dominant contributions to the inverse Laplace transform are geometries with one shell inside the two horizons of the double-sided black hole with masses $E_L$ and $E_\i$, while the second shell of mass $m$ is outside the right horizon $r_\i$ at $t=0$. The latter shell was emitted from the past horizon of the right black hole and will eventually fall back into the future horizon, increasing the mass $E_i$ to $E_R$.\footnote{Restricting to $E_R>E_\i$ is a necessary condition for the shell to be able to be emitted from the right black hole. If $E_R<E_\i$, the fixed mass $m$ shell will instead be inside of the right horizon, and the corresponding microstates are not consistent with a large fluctuation leaving the horizon, see \autoref{fig:externalshell2} (b). In this case, the universal results of~\cite{shells1} and section~\ref{sec: large mass} imply that the dimension spanned by such states is $e^{\bf{S}_L+\bf{S}_R}$.} Importantly, the difference $E_R-E_\i$ carried by the shell of matter is independent of its inertial mass $m$. In the small mass regime $m\ll m_0$, we are considering an ultra-relativistic limit in which most of the energy of the shell is kinetic. Taking the strict $m \to 0$ limit corresponds to a shell whose energy is composed purely of radiation so that the emitted shell reaches the asymptotic boundary and bounces back to the black hole. 

\paragraph{Moments of overlaps.}
The connected moments of overlaps~\eqref{eq: nth overlap two} can be computed analogously to the procedure outlined in~\autoref{sec: microoverlaps}. Specifically, we will compute the numerator and the denominator of~\eqref{eq: nth overlap two} using the gravity path integral. 

The term in the denominator of~\eqref{eq: nth overlap two} is dominated by disconnected contributions. To this end, we focus on computing $\overline{\bcZ_{kk}}$, which is defined through the on-shell actions that normalize the microcanonically projected states in~\eqref{eq: microstate two shells} in the small mass $m$ limit. Hence, we need to find the classically allowed geometries filling in the boundary conditions in~\autoref{fig:contour}(b) and compute their on-shell action. When the masses of the operators $\mO^{(i)\dagger}$ and $\mO^{(j)}$ are equal, $m_i=m_j=i\mu$ where $\mu$ is taken to be large, this heavy shell is inside both horizons, and the geometry is given in~\autoref{fig:externalshell2}. The temperatures in the different sections of the geometry are given by
\begin{equation}\label{eq: constrains small m}
    \Bar{\beta}_L=\beta_L\,, \quad \frac{\Bar{\beta}_\i+\Bar{\beta}_\i'}{2} = \beta_\i-\frac{\beta_\i}{\pi}\arcsin\left(\frac{r_\i}{R_*}\right)
    \,, \quad \Bar{\beta}_R= \frac{\beta_R}{\pi}\arcsin\left(\frac{r_R}{R_*}\right)\,,
\end{equation}
where the turning point of the shell with mass $m$ is given by 
\begin{equation}\label{eq: Rstar}
    R_*^{2}=r_R^{2}+\left(\frac{M_R-M_\i}{m}-2G m\right)^2\,.    
\end{equation}
The on-shell action that computes $\overline{\cZ_{kk}\left(\bar{\beta}_\i,\bar{\beta}'_\i\right)}$ is diagrammatically given by
\be
I_{\rm tot}^{\rm ren}\Big|_{\rm on-shell}=\IPL{\,L}^{\rm ren}+\ICL{\,L}^{\rm ren}+I_{\textcolor{red}{\cal W}}^{\rm ren} +\ICR{\,\i}^{\rm ren}++\IPiR{\,\i}^{\rm ren}+\IPiR{\,\i'}^{\rm ren}+\ICL{\,\i}^{\rm ren}+I_{\textcolor{purple}{\cal W}}^{\rm ren} -\ICL{\,R}^{\rm ren} +\IPiR{\,R}^{\rm ren}\,,
\ee
where the various contributions are given in \autoref{sec: building blocks}. In the pinching limit $\mu\to\infty$, we find
\begin{equation}
\begin{aligned}
    -\log  \overline{\cZ_{kk}} =& \Bar{\beta}_L \mathcal{F}(\Bar{\beta}_L) + 2\mu \log \frac{1}{G \mu} +\frac{\bar{\beta}_\i+\bar{\beta}_\i'}{2} \mathcal{F}(\beta_\i) \\
    &+ \frac{\sqrt{R_*^2-r_\i^2}-\sqrt{R_*^2-r_R^2}}{2G} \log\frac{2}{R_*}+\bar{\beta}_R \mathcal{F}(\beta_R)\,.
\end{aligned}
\end{equation}

In the small $m$ limit, the relation between the Euclidean lengths $\bar{\beta}_\i$, $\bar{\beta}'_\i$, $\bar{\beta}_R$ and the black hole temperatures $\beta_\i$ and $\beta_R$ is
\begin{equation}
   \beta_\i = \frac{\bar{\beta}_\i+\bar{\beta}_\i'}{2} +\bar{\beta}_R  +\mathcal{O}\left(m^{3}\right)\,, \quad \beta_R = \frac{  \Bar{\beta}_\i + \Bar{\beta}_\i' }{2\sqrt{1 +\frac{ (\Bar{\beta}_\i+\Bar{\beta}_\i')^2 G m}{\pi^2\bar{\beta}_R}}}+\frac{ \Bar{\beta}_R  }{\left(1 +\frac{(\Bar{\beta}_\i+\Bar{\beta}_\i')^2 G m}{\pi^2\bar{\beta}_R}\right)^{3/2}}+\mathcal{O}\left(m^{2}\right)\,.
\end{equation}
Importantly, note that $R_* \approx \frac{M_R-M_\i}{m}$ in the small mass limit (see eq.~\eqref{eq: Rstar}), so from eq.~\eqref{eq: constrains small m}, it follows that $\bar{\beta}_R$ is linear in $m$. The on-shell action of the geometry computing the norm of the state in this regime is therefore
\begin{equation}\label{eq: onshell action two shells small mass norm}
\begin{aligned}
    -\log  \overline{\cZ_{kk}} = & \bar{\beta}_L \mathcal{F}(\bar{\beta}_L)+\frac{\bar{\beta}_\i+\bar{\beta}_\i'}{2}  \mathcal{F}\left(\frac{\bar{\beta}_\i+\bar{\beta}_\i'}{2} \right) +2 \mu \log \frac{1}{G \mu}\,\\
    &+\frac{\bar{\beta}_R}{16 G}+2 m \log (\bar{\beta}_R)-2 m+\bar{\beta}_R \bar{M}_\i-\frac{4\pi ^2 \bar{\beta}_R^2 }{ \left(\bar{\beta}_\i+\bar{\beta}'_\i\right)^3 G }+\mathcal{O}\left(m^{3}\right) \,,
\end{aligned}
\end{equation}
where $\overline{M}_\i = \frac{2\pi^2}{G (\bar{\beta}_\i+\bar{\beta}'_\i)^2}$.

The saddle point values of the Euclidean lengths are given by\footnote{In standard statistical mechanics, there is an equivalence between microcanonical and canonical ensembles in the thermodynamic limit. One might worry that such an equivalence breaks down for atypical fluctuations. Therefore, a prior, the saddle point approximation for the inverse Laplace transform may not be valid -- we thank Alejandro Vilar L\'opez for raising this point. However, we verified that this approximation remains valid by computing the eigenvalues of the Hessian at the saddle point and confirming that the distribution is sharply peaked at the saddle point value.}
\begin{equation}
    \bar{\beta}_L = \frac{\pi^2}{G  \bS_L} \,,\quad \frac{\bar{\beta}_\i+\bar{\beta}_\i'}{2}  = \frac{\pi^2}{G \bS_\i} -\frac{2m }{E_R-E_\i}+\mathcal{O}\left(m^{3}\right)\,,
    \quad \bar{\beta}_R = \frac{2m}{E_R-E_\i} +\mathcal{O}\left(m^{3}\right)\,,
\end{equation}
where the entropies $\bS_{L,\i}$ are related to the energies $E_{L,\i}$ by eq.~\eqref{eq: entropy}. Therefore, the saddle point approximation of the normalization constant is
\begin{equation}
\overline{\bcZ_{kk} } = e^{-\bI_1}\,,
\end{equation}
to leading order in the semiclassical approximation, where
\begin{equation}
    \bI_1 = \bS_L+\bS_\i+2 \mu  \log \frac{1}{G \mu}+4m-2 m \log \left(\frac{2m}{E_R-E_\i}\right)+\mathcal{O}\left(m^{3}\right) \,.
\end{equation} 
The term in the denominator of~\eqref{eq: nth overlap two} is dominated by the fully disconnected contribution, and the corresponding semiclassical approximation is
\begin{equation} \label{eq: denominator}
    \overline{\bcZ_{k_1 k_1}\bcZ_{k_2 k_2} \ldots \bcZ_{k_{n} k_{n}}} = e^{-n \bI_1}\,.
\end{equation}

\begin{figure}[t]
    \centering
    \includegraphics[scale=0.65]{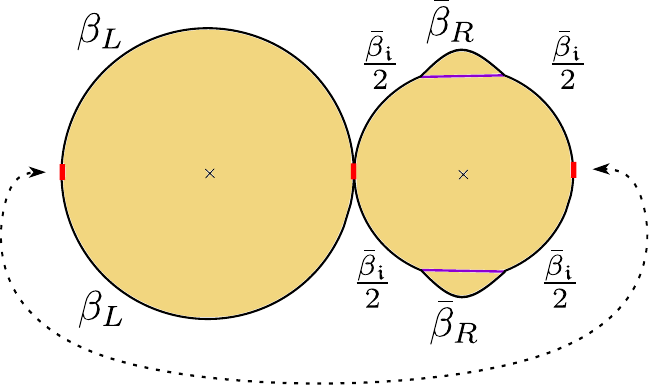} \quad 
    \includegraphics[scale=0.65]{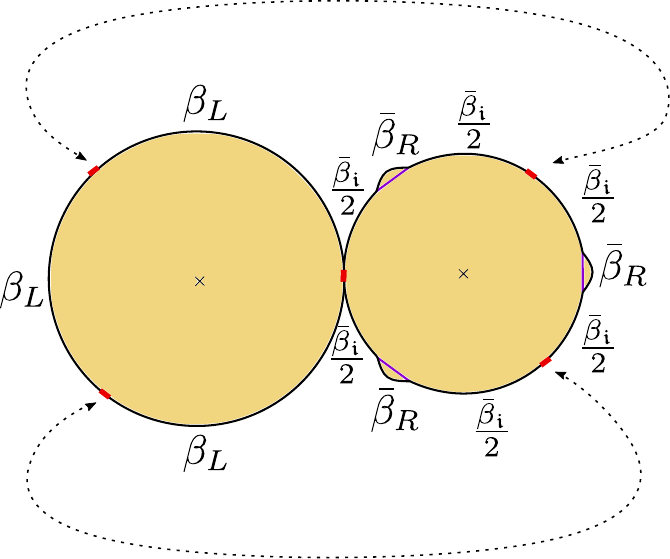} \\ 
    (a) \hspace{5cm} (b)
    \caption{Wormhole geometries which dominate the connected contributions to moments of overlaps of the states in~\eqref{eq: microstate two shells} with small mass $m$ when $E_R$ is greater than $E_{\i}$. (a) Two-boundary wormhole which computes the connected contribution to $\overline{\cZ_{ij}\cZ_{ji}}$. (b) Three-boundary wormhole which computes the connected contribution to $\overline{\cZ_{ik}\cZ_{kj}\cZ_{ji}}$. Both geometries are given in the pinching limit $\mu\to\infty$ for which $\Bar{\beta}_L\to\beta_L$.} 
    \label{fig:butterfly}
\end{figure}

We now proceed to compute the numerator of~\eqref{eq: nth overlap two}. To this end, we begin by computing $ \left.\overline{\cZ_{k_1 k_2}\cZ_{k_2 k_3}\ldots \cZ_{k_{n} k_1}}\right|_{\rm conn.}$. The relevant geometries for $n=2$ and $n=3$ in the limit where $\mu\to\infty$ are shown in~\autoref{fig:butterfly}.
The temperatures of the various regions are related to the Euclidean preparation times by
\begin{equation}
    n q_L=\tilde{\beta}_L\,,\quad n\frac{q_\i+q'_\i}{2}= \Tilde{\beta}_\i-\sum_i\frac{\Tilde{\beta}_\i}{\pi}\arcsin\left(\frac{\Tilde{r}_\i}{\Tilde{R}_*^{(i)}}\right)\,,\quad  \Bar{\beta}_R^{(i)}=\frac{\Tilde{\beta}_R^{(i)}}{\pi}\arcsin\left(\frac{\Tilde{r}^{(i)}_R}{\Tilde{R}_*^{(i)}}\right)\,,
    \end{equation}
where $q_{L,R} = \frac{1}{n}\sum_i \Bar{\beta}^{(i)}_{L,R}$, $q_{\i} = \frac{1}{n}\sum_i \Bar{\beta}^{(i)}_{\i}$, and $q'_{\i} = \frac{1}{n}\sum_i \Bar{\beta}^{'(i)}_{\i}$. The turning points of the shells of mass $m$ are all at the same location
\begin{equation}
\Tilde{R}_*^{(i)2}=\Tilde{r}_R^{(i)2}+\left(\frac{\Tilde{M}^{(i)}_R-\Tilde{M}_\i}{m}-2G m\right)^2\,,
\end{equation}
where $\tilde{M}_\i = \frac{\pi^2}{2G \Tilde{\beta}_\i^2}$ and $\tilde{M}^{(i)}_R = \frac{\pi^2}{2G \Tilde{\beta}_R^{(i)2}}$. The corresponding on-shell action leads to 
\begin{equation}
\begin{aligned}
    - \log  \left.\overline{\cZ_{k_1 k_2}\ldots \cZ_{k_{n} k_1}}\right|_{\rm conn.} = & n q_L \mathcal{F}(\tilde{\beta}_L) +n \frac{q_\i + q_\i'}{2} \mathcal{F}(\tilde{\beta}_\i) + \sum_i^n \bar{\beta}^{(i)}_R \mathcal{F}(\tilde{\beta}^{(i)}_R) \\
    & + \sum_i^n\frac{\sqrt{R_*^{(i)2}-\tilde{r}_\i^2}-\sqrt{R_*^{(i)2}-\tilde{r}_R^{(i)2}}}{2G} \log\frac{2}{R_*^{(i)}}+2 \mu n \log \frac{1}{G \mu} \,.
\end{aligned}
\end{equation}

In the limit of small mass $m$,
the relation between the average Euclidean lengths $q_\i$, $q_\i'$ and $q_{L,R}$ and the temperatures $\tilde{\beta}^{(i)}_R$ and $\tilde{\beta}_{L,\i}$ is
\begin{equation}
\begin{aligned}
   \tilde{\beta}_L & = 
 n q_L\,,\hspace{3 cm} \tilde{\beta}_\i = n\frac{q_\i+q_\i'}{2} + n q_R +\mathcal{O}\left(m^{3}\right)\,, \\ \tilde{\beta}_R^{(i)} & = \frac{n}{2}\frac{  q_\i + q_\i' }{\sqrt{1 +\frac{ n^2(q_\i+q_\i')^2 G m}{\pi^2 \bar{\beta}_R^{(i)}}}}+\frac{ n q_R  }{\left(1 +\frac{ n^2(q_\i+q_\i')^2 G m}{\pi^2 \bar{\beta}_R^{(i)}}\right)^{3/2}}+\mathcal{O}\left(m^{2}\right)\,.
 \end{aligned}
\end{equation}
In the small $m$ regime, the on-shell action of the geometry computing $\left.\overline{\cZ_{k_1 k_2}\ldots \cZ_{k_{n} k_1}}\right|_{\rm conn.} $ is
\begin{equation}
\begin{aligned}
    - \log  \left.\overline{\cZ_{k_1 k_2}\ldots \cZ_{k_{n} k_1}}\right|_{\rm conn.} = &\ n q_L \mathcal{F}(n q_L )+n\frac{q_\i+q_\i'}{2}  \mathcal{F}\left(n\frac{q_\i+q_\i'}{2} \right) +2 n \mu \log \frac{1}{G \mu}\,\\
    &+\frac{n q_R}{16 G}+\sum_i 2 m \log (\bar{\beta}_R^{(i)})-2n m+\frac{2\pi^2q_R}{Gn (q_\i+q'_\i)^2}\\
    &-  \frac{4\pi^2q_R^2 }{G n (q_\i+q'_\i)^3}+\mathcal{O}\left(m^{3}\right) \,.
\end{aligned}
\end{equation}
The saddle point values of the Euclidean lengths are given by
\begin{equation}
    q_L = \frac{\pi^2}{G n \bS_L} \,,\quad   
    \quad \frac{q_\i+q_\i'}{2}  = \frac{\pi^2}{G \bS_\i} -\frac{2m }{E_R-E_\i}+\mathcal{O}\left(m^{3}\right)\,,
    \quad q_R = \frac{2m}{E_R-E_\i} +\mathcal{O}\left(m^{3}\right)\,.
\end{equation}
Therefore, the saddle point approximation to the term in the numerator of~\eqref{eq: nth overlap two} is
\begin{equation}\label{eq: numerator}
\left.\overline{\bcZ_{k_1 k_2}\ldots \bcZ_{k_{n} k_1}}\right|_{\rm conn.} = e^{-\bI_n}\,,
\end{equation}
to leading order in the semiclassical approximation. In the small $m$ regime one finds
\begin{equation}
    \bI_n= \bS_L+\bS_\i+2 n \mu  \log \frac{1}{G \mu}+4nm-2 n m \log \left(\frac{2m}{E_R-E_\i}\right)+\mathcal{O}\left(m^{3}\right)  \,.
\end{equation}

Using the results~\eqref{eq: denominator} and \eqref{eq: numerator} in~\eqref{eq: nth overlap two}, we find the semiclassical approximation to the connected component of the $n$-moment of overlaps is\footnote{While we explicitly display the derivation up to ${\cal O}(m^{3})$, we have verified it up to ${\cal O}(m^{6})$. However, unlike in the large mass regime of~\autoref{sec: large mass}, we do not have an argument for why the corrections should vanish. In fact, as will be noted at the end of this section, there are non-perturbative corrections to the dimension.}
\be
\left.\overline{\langle \psi^E_{k_1} | \psi^E_{k_2} \rangle \langle \psi^E_{k_2} | \psi^E_{k_3} \rangle \ldots \langle \psi^E_{k_{n}} | \psi^E_{k_1} \rangle} \right|_{\text{conn.}} =e^{ (1-n) \left(\bS_L+\bS_\i\right)+\mathcal{O}\left(m^{3}\right)}\,. \label{eq: microoverlap-n ext}
\ee

\paragraph{Hilbert space dimension.} When $E_R$ is greater than $E_{ \i}$ and the second shell has a small mass, the connected contribution to the moments of overlaps~\eqref{eq: microoverlap-n ext} is the same as in the single shell states~\eqref{eq: microoverlap-n} up to $ \mathcal{O}\left(m^{3}\right)$ but with $\bS_R$ replaced by $\bS_\i$. This implies that the Gram matrix of the states in~\eqref{eq:Kspanmicro2} in the small mass limit is equal to the one in~\autoref{sec:review} up to $ \mathcal{O}\left(m^{3}\right)$ but with the replacement $\bS_R\to \bS_\i$.  So the results of \autoref{sec: the resolvent} apply to the Gram matrix of states with a second shell with a fixed but small mass, and in particular, the rank of the matrix is
\begin{equation}\label{eq: Rank small m}
{\rm Rank} \left(G_{ij}\right) = 
    \begin{cases}
        \Omega\quad \quad \quad {\rm for}  \quad \Omega \leq e^{\bS_L+\bS_\i} \,, \\
        e^{\bS_L+\bS_\i} \quad \ {\rm for} \quad \Omega \geq e^{\bS_L+\bS_\i} \,,
    \end{cases}
\end{equation}
up to possible corrections for finite $m$, see~\autoref{foot: stability}. The maximal rank of the  Gram matrix corresponds to the Hilbert space dimension spanned by these atypical states. As expected, it is a much smaller subspace of the Hilbert space of the microcanonical double-sided black hole since $\bS_\i<\bS_R$ because $E_\i<E_R$. This agrees with the expectations from the analogous question in statistical mechanics of how many microscopic states are consistent with some atypical fluctuations. The prototypical example is a room filled with a gas of particles where a significant fraction is concentrated in a corner. 

To conclude, notice that the dimension spanned by the states with shells of small mass~\eqref{eq: Rank small m} is perturbatively independent on $m$ up to the order we have verified and saturates the bound on the dimension coming from the counting argument presented below \eqref{eq:Kspanmicro2}. From the boundary perspective, the saturation of the bound is tied to the ETH ansatz for the operators ${\cal O}^{(k)}$, and more precisely, it relies on its energy basis coefficients being randomized coefficients. This amounts to ${\cal O}^{(k)}_{mn}$ being uncorrelated random matrices, and thus, the upper bounds argued below~\eqref{eq:Kspanmicro2} should be saturated as $\Omega$ becomes large enough. The situation is more subtle without resorting to ETH or when viewing the problem from the bulk perspective. Intuitively, the bound saturation may have been expected in the cases where the intermediate energy is large $(E_\i>E_R)$ or the mass is large $(m>m_0)$  since the shells are hidden behind the horizon. However, for small masses $(m<m_0)$ and small intermediate energy $(E_\i<E_R)$, the second shell of matter is not behind the right horizon, and therefore, a similar intuitive explanation for why the bound on the dimension should be saturated may not hold. The question of whether higher order perturbative or even non-perturbative corrections could affect the dimension of the Hilbert space is left for future work.

\section{Discussion}
\label{sec:discussion}
In this paper, we construct a family of semiclassical black hole microstates consistent with an atypical fluctuation where an order one fraction of the energy of an eternal black hole is emitted in a shell of matter or radiation coming from the past horizon and re-absorbed later at the future horizon. We then use the gravitational path integral to find the dimension of the corresponding Hilbert space. We do this by adding a second shell to the framework of single-shell semiclassical black hole microstates in \cite{shells1}.  In our state construction, if the intermediate energy $E_\i$ is larger than the right energy $E_R$, the second shell is always behind the horizon and thus leads to the same external geometry.  In this regime, the presence of a second shell of matter does not change the spanned Hilbert space, as expected from the arguments in  \cite{shells1}.  On the other hand, if the intermediate energy $E_\i$ is lower than $E_R$, we find that the microstates we construct span a Hilbert space whose dimension is set by the horizon area of the intermediate black hole instead of the right black hole. If the second shell is heavy $(m>m_0)$ and therefore is behind the horizons, this is somewhat surprising and highlights a subtlety in the arguments of~\cite{shells1}. Nevertheless, the reduced dimension of the Hilbert space agrees with the counting argument presented below~\eqref{eq:Kspanmicro2}. Moreover, if the second shell is sufficiently light, it will exit the past horizon and re-enter the future horizon.  In this case, the reduction of the Hilbert space dimension is expected from physical arguments -- the external observer can measure everything about the emitted shell, and so it does not contribute in any way to the external observer's ignorance of the microstate, and, hence the entropy.
 
Specifically, our calculation regarding the out-of-equilibrium fluctuations of black holes focuses on microstates for which the additional shell has very small mass $m\ll m_0$, and increases the energy from $E_\i$ to $E_R$. In this case, the Lorentzian trajectory of the additional shell starts at the white hole singularity in the far past. After some time, it is emitted through the past horizon along a trajectory that is almost null, after which it turns around at the $t=0$ slice to be later reabsorbed through the future horizon of the black hole until it eventually reaches the black hole singularity as shown in \autoref{fig:shelloutside} (c). These two-shell microstates correspond to double-sided (microcanonical) black holes with different temperatures (energies), and the atypical fluctuation occurs on the right black hole, where we focus our discussion. In the canonical states, these solutions correspond to a black hole at temperature $\beta_R^{-1}$, which emits a large amount of energy in the form of a spherical shell of mass $m$, lowering its temperature transiently to $\beta_\i^{-1}$. Because of the AdS gravitational potential, the spherical shell of matter turns back toward the black hole and falls in, reheating it back to its initial temperature, $\beta_R^{-1}$. In the microcanonical states, the initial microcanonical black hole is at energy $E_R$, which is lowered after the emission of the shell to energy $E_\i$.  After reabsorbing the shell, the energy of the microcanonical black hole returns to $E_R$. 

Most of our explicit calculations were carried out in AdS$_3$, where analytic calculations are possible, and in situations where the second shell was very heavy or very light. However, we expect that our results will generalize to higher dimensions, and to situations where the second shell has an intermediate mass.  First, consider our state construction in any dimension, with a second shell with arbitrary positive mass that separates the two regions with fixed energies $E_\i$ and $E_R$ where $E_\i >E_R$. In this setup, both shells are always behind the horizon, and the two-shell states should still precisely span the black hole microstate Hilbert space in the same way as in AdS$_3$. Next, consider the case where $E_R>E_\i$.  When the second shell has a large mass $(m>m_0)$, by a simple counting argument we expect that these states span a subspace of the black hole Hilbert space. When the second shell has a small mass $(m<m_0)$, this is by construction a microstate of an atypical fluctuation because the shell will exit and then re-enter the black hole. In this case, we showed above that the spanned Hilbert has a lower dimension as expected, but additionally, the computations were perturbatively independent of the inertial mass $m$ of the second shell up to order $\mO(m^6)$. The question of whether this holds in general dimension, or even in AdS$_3$ to higher order in $m$ is left for future investigation. Additionally, these corrections should preserve the fact that the dimension of the Hilbert space is a natural number. It would be useful to explore this further.

Note that the microcanonical states that we worked with can be obtained from the canonical states by an inverse Laplace transform $|\Psi^E \rangle \sim \int d\bar{\beta}_L d\bar{\beta}_L e^{\bar{\beta}_L E_L/2 +\bar{\beta}_R E_R/2} \sqrt{Z_\Psi} |\Psi \rangle$ -- see also appendix B of~\cite{Balasubramanian:2024yxk}. Because of the integral over temperatures, these states are linear superpositions of canonical states in many configurations: both typical and atypical, as well as ones for which the shell is internal, and ones for which the shell is emitted on the left or right. This resonates with the idea that atypical fluctuations are expected to occur within the Hilbert space of typical states in analogy to the statistical picture. The choice of having small mass $m< m_0$ and $E_R>E_\i$ ensures that these atypical states dominate the inverse Laplace transform: the microcanonical states with $m < m_0$ are superpositions of mostly atypical canonical states. 

In this paper, we considered time-symmetric fluctuations of black holes. However, an arbitrary linear combination of these states can break time-reversal symmetry.\footnote{This can be explicitly worked out for coherent states of a harmonic oscillator, where a linear combination of coherent states with zero momentum can give a coherent state with non-zero momentum.}  This might be a way to construct microstates for other dynamical black hole fluctuations. Examples of this are black hole formation from collapse of matter, and evaporating black holes. Fleshing out the statistical interpretation of the Bekenstein-Hawking entropy of such black holes would be interesting to do. From a broader perspective, this work can be considered a step towards a better understanding of the out-of-equilibrium thermodynamics of black holes.

\paragraph{Acknowledgements: } We would like to thank Will Chan, Shira Chapman, Charlie Cummings, Chitraang Murdia, Martin Sasieta, Alejandro Vilar L\'opez, and Tom Yildirim for useful discussions.  VB was supported in part by the DOE through DE-SC0013528 and QuantISED grant DE-SC0020360, and in part by the Eastman Professorship at Balliol College, University of Oxford. Work at VUB was supported by FWO-Vlaanderen project G012222N and by the VUB Research Council through the Strategic Research Program High-Energy Physics. JH is supported by FWO-Vlaanderen through a Junior Postdoctoral Fellowship. Mikhail Khramtsov would like to thank the Steklov Mathematical Institute of the Russian Academy of Sciences for hospitality during the final stages of the work.

\appendix
\section{Thin shell trajectories in AdS$_3$}
\label{app: D=3}

In this appendix, we give the explicit equations for the shell trajectories described in~\autoref{sec: shells} for the case of $d=2$. We start by solving the equations of motion of the shell in Euclidean signature, which are given by~\eqref{shell-EOM-1} and~\eqref{shell-EOM-2}.
The equations are solved by\footnote{Note that when the value of $R_*$ approaches $r_+$, the $\tau_+$ trajectory becomes simply
\begin{equation}
    \lim_{R_*\to r_+}\tau_+(T)= \tau_+^*+\frac{\beta_+}{4}{\rm sign}(T)\,.
\end{equation}
}
\be\label{eq: trajectories}
\begin{aligned}
R(T) &= R_* \cosh T\,,\\
\tau_\pm(T) &=\tau^*_\pm+\frac{\beta_\pm}{2\pi} \arctan\left(\frac{r_\pm \tanh(T)}{\sqrt{R_*^2-r_\pm^2}} \right)\,,
\end{aligned}
\ee
where $\tau_\pm^* = \tau_\pm(0)$ and we recall that $
R_* = \sqrt{r_+^2 + \left(\frac{M_+ - M_-}{m} -2 G m\right)^2}$. We have chosen the initial conditions such that the shell passes through the turning point $R_*$ at $T=0$. Notice that at the turning point, the elapsed Euclidean time $\tau_\pm^*$ is fixed by the initial conditions as explained in~\autoref{sec: shells} such that $\tau_\pm^*$ is either $\beta_\pm / 2$ or $0$. Then the integral in eq.~\eqref{DeltaTau} is given by
\be
\Delta \tau_\pm = \frac{\beta_\pm}{\pi} \arcsin\left(\frac{r_\pm}{R_*}\right)\,.
\ee
We can use eq.~\eqref{eq: trajectories} to relate $R$ and $\tau_\pm$ as
\begin{equation}\label{eq: R traj euclidean}
    R(\tau_\pm) = \frac{R_*}{\sqrt{1-\alpha^2}}\,,
\end{equation}
where
\begin{equation}
    \alpha^2 = \frac{R_*^2-r_\pm^2}{r_\pm^2} \tan^2 \left( \frac{2\pi \tau_\pm}{\beta_\pm}\right)\,.
\end{equation}

Next, we can analytically continue the trajectories in eq.~\eqref{eq: R traj euclidean} to Lorentzian signature using $\tau_\pm \to i t_\pm$, and
\begin{equation}\label{eq: R traj lorentzian}
    R(t_\pm) = \frac{R_*}{\sqrt{1+\gamma^2}}\,,
\end{equation}
where
\begin{equation}
    \gamma^2 = \frac{R_*^2-r_\pm^2}{r_\pm^2} \tanh^2 \left( \frac{2\pi t_\pm}{\beta_\pm}\right)\,. 
\end{equation}
These trajectories are time-reversal symmetric, begin at $R=R_*$ for $t_\pm=0$ and fall to the future (past) event horizon for $t_\pm \to \infty$ ($t_\pm \to -\infty$). 

To continue these trajectories past the horizons $r_\pm$, we define the infalling coordinates
\begin{equation}
    v_\pm = t_\pm -\int^\infty_R \frac{dr'}{f_\pm(r')} = t_\pm - \frac{1}{r_\pm}{\rm arcoth}\left(\frac{R}{r_\pm}\right)\,, \quad {\rm for} \quad  R>r_\pm\,,
\end{equation}
for which the trajectories are

\begin{equation}
\label{eq: trajectories-L}
\begin{aligned}
    R(S) & = R_* \cos(S)\,,\\
    v_\pm (S) &= \frac{\beta_\pm}{2\pi }{\rm arctanh}\left(\frac{r_\pm \tan (S)}{\sqrt{R_*^2-r_\pm^2}}\right) - \frac{\beta_\pm}{2\pi }{\rm arcoth}\left(\frac{R_* \cos(S)}{r_\pm}\right)\,,
\end{aligned}
\end{equation} where we have analytically continued the Euclidean parameter $T$ for the trajectories in~\eqref{eq: trajectories} to $T\to iS$ as well.
We can also parametrize the paths in terms of the radius, which gives 
\begin{equation}
\begin{aligned}
    v_\pm(R) &= \frac{\beta_\pm}{2\pi} \left({\rm arctanh}\left(\frac{r_\pm}{R} \frac{\sqrt{R_*^2-R^2}}{\sqrt{R_*^2-r_\pm^2}}  \right)- {\rm arctanh} \left(\frac{r_\pm}{R}\right) \right)\,,\quad  {\rm for} \quad R>r_\pm\,.
\end{aligned}
\end{equation}
The infalling coordinates remain finite throughout the trajectory and cross the horizon at $v_\pm = \frac{1}{2r_\pm} \log \left(\frac{R_*^2-r_\pm^2}{R_*^2} \right)$. Furthermore, we can define an inside Schwarzschild time $\Tilde{t}_\pm$ via
\begin{equation}
    v_\pm = \Tilde{t}_\pm + \int^R_0 \frac{dr'}{f_\pm(r')} = \Tilde{t}_\pm - \frac{1}{r_\pm}{\rm arctanh}\left(\frac{R}{r_\pm}\right)\,, \quad {\rm for} \quad  R<r_\pm\,.
\end{equation}
The trajectory inside the horizon is
\begin{equation}\label{eq: trajectories-L2}
\begin{aligned}
    R(S) & = R_* \cos(S)\,,\\
     v_\pm (S) &= \frac{\beta_\pm}{2\pi } \left({\rm arctanh}\left(\frac{\sqrt{R_*^2-r_\pm^2}}{r_\pm \tan (S)}\right) - {\rm arctanh}\left(\frac{R_* \cos(S)}{r_\pm}\right) \right)\,,
\end{aligned}
\end{equation}
or, in terms of the radius $R$,
\begin{equation}
\begin{aligned}
    v_\pm(R) &= \frac{\beta_\pm}{2\pi} \left({\rm arctanh}\left(\frac{R}{r_\pm} \frac{\sqrt{R_*^2-r_\pm^2}}{\sqrt{R_*^2-R^2}}  \right) - {\rm arctanh} \left(\frac{R}{r_\pm}\right) \right)\,,\quad  {\rm for} \quad R<r_\pm\,.
\end{aligned}
\end{equation}

To conclude, the trajectory of the shell is given by eqs.~\eqref{eq: trajectories} in the Euclidean geometry, and by eqs.~\eqref{eq: trajectories-L} and~\eqref{eq: trajectories-L2} in the Lorentzian geometry for the case $d=2$.

\section{Orientation prescription for gravity saddles}
\label{sec: orientation}

In this appendix, we clarify the orientation rules and conventions used in applying the saddle point approximation to the gravitational path integral for computing inner products of states containing shells, denoted by an overline. These inner products are expressed using Euclidean time contours. For example, the overlap of two states, each containing a single shell insertion, is depicted as follows
\be
\langle \Psi_i | \Psi_j \rangle = \includegraphics[scale=0.35,valign = c]{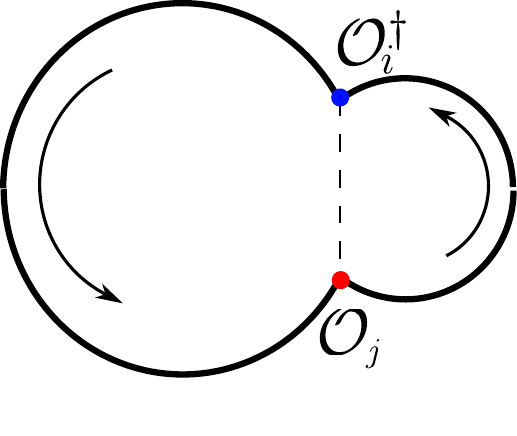}\,. \label{eq: overlap-fig}
\ee
We will always keep ket states on the bottom half of the contour and bra states on the top half of the contour. Note that the flow of Euclidean time along the contour, shown by the arrow, equips the time contour with a natural orientation. 

Generally speaking, we are interested in computing
\be
\overline{\langle \Psi_{j_1} | \Psi_{j_2} \rangle \langle \Psi_{j_2} | \Psi_{j_3} \rangle \dots \langle \Psi_{j_{n}} | \Psi_{j_{1}} \rangle}\,. \label{eq: generic overlap}
\ee
For the purposes of this appendix, the states $|\Psi_{j_k} \rangle$ can contain any number of shells inside or outside the horizon. When using the saddle point approximation to the gravity path integral to find \eqref{eq: generic overlap}, one constructs all Euclidean spacetime geometries that fill the specified contours. The contours themselves, when combined with suppressed spatial dimensions, act as the asymptotic boundaries of the spacetime geometry. An important contribution to such overlaps comes from the most connected geometry. This geometry is a Euclidean wormhole with $n$ asymptotic boundaries. To construct it, one takes $n$ contours, one for each inner product of the form \eqref{eq: overlap-fig}. Then, one needs to connect the matching operator insertions with shell worldlines. Finally, one fills in the spacetime between pairs of shell worldlines with segments of Euclidean AdS spacetime, ensuring that the resulting geometry satisfies the Israel junction conditions.

In general, a wormhole constructed in such a way is not unique. However, the precise ordering of operators within the overlaps in~\eqref{eq: generic overlap}  combined with the orientation induced by the flow of Euclidean time on the contours, imposes strong constraints ensuring that the wormhole geometry is unique. 
\begin{figure}
	\centering
	\includegraphics[scale=0.4]{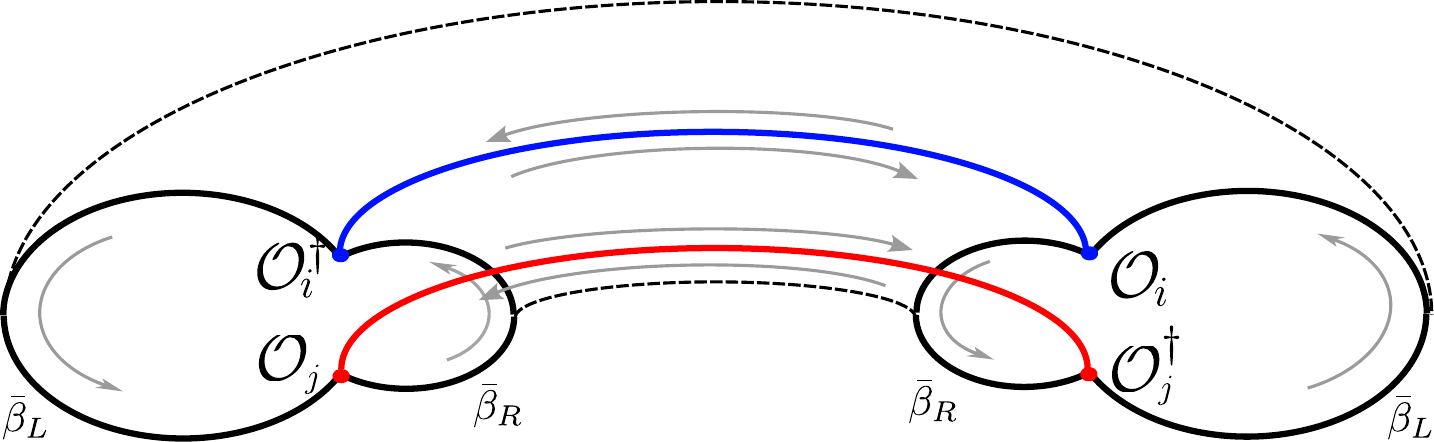}\qquad
	\includegraphics[scale=0.5]{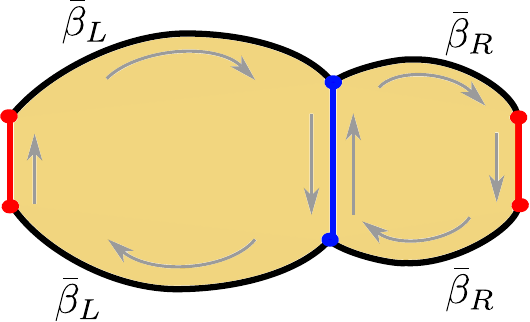}\\
	(a) \hspace{7cm} (b)
	\caption{Two-boundary wormhole computing the connected contribution to $\overline{\langle \Psi_i | \Psi_j \rangle \langle \Psi_j | \Psi_i \rangle }$ (no sum over indices). Red and blue lines are the shell worldvolumes. The gray arrows denote the orientation of the asymptotic boundary segments and shell worldlines, induced by the flow of Euclidean time in each of the two segments. Panel (a) shows the topology, where the Euclidean contour computing $\langle \Psi_j | \Psi_i \rangle$ has been rotated by $\pi$ rad for convenience. Panel (b) schematically shows the gluing of the two Euclidean AdS black hole segments along the shell worldlines (the two copies of the red worldline are identified). }
	\label{fig: overlap-magnitude-squared}
\end{figure}
To illustrate this, consider the simple example of constructing the two-boundary wormhole geometry for $\overline{\langle \Psi_i | \Psi_j \rangle \langle \Psi_j | \Psi_i \rangle}$, where each state has a single operator insertion and the indices are not summed over. To this end, one takes two copies of the Euclidean time contour, one computing the overlap $\langle \Psi_i | \Psi_j \rangle$ and another computing $\langle \Psi_j | \Psi_i \rangle$. The matching operators on these contours are connected using shell worldlines, as shown in~\autoref{fig: overlap-magnitude-squared}(a). The resulting spacetime consists of two regions of the Euclidean AdS black hole solution, glued along the shell worldlines via Israel junction conditions, as shown in~\autoref{fig: overlap-magnitude-squared}(b). A critical aspect of this construction is ensuring that the orientation of the asymptotic boundary segments, as induced by the flow of Euclidean time on the original overlap contours, remains continuous throughout the wormhole solution. In ~\autoref{fig: overlap-magnitude-squared}(b), this is reflected by the fact that the gray arrow is cyclic in each of the two segments. This condition uniquely determines the wormhole solution.

\bibliographystyle{JHEP}
\bibliography{shells.bib}

\providecommand{\href}[2]{#2}\begingroup\raggedright\begin{thebibliography}{10}

\bibitem{shells1}
V.~Balasubramanian, A.~Lawrence, J.M.~Magan and M.~Sasieta, \emph{{Microscopic
  origin of the entropy of black holes in general relativity}},
  \href{https://arxiv.org/abs/2212.02447}{{\ttfamily 2212.02447}}.

\bibitem{shells2}
V.~Balasubramanian, A.~Lawrence, J.M.~Magan and M.~Sasieta, \emph{{Microscopic
  Origin of the Entropy of Astrophysical Black Holes}},
  \href{https://doi.org/10.1103/PhysRevLett.132.141501}{\emph{Phys. Rev. Lett.}
  {\bfseries 132} (2024) 141501}
  [\href{https://arxiv.org/abs/2212.08623}{{\ttfamily 2212.08623}}].

\bibitem{Climent24}
A.~Climent, R.~Emparan, J.M.~Magan, M.~Sasieta and A.~Vilar~L\'opez,
  \emph{{Universal construction of black hole microstates}},
  \href{https://doi.org/10.1103/PhysRevD.109.086024}{\emph{Phys. Rev. D}
  {\bfseries 109} (2024) 086024}
  [\href{https://arxiv.org/abs/2401.08775}{{\ttfamily 2401.08775}}].

\bibitem{Balasubramanian:2024yxk}
V.~Balasubramanian, B.~Craps, J.~Hernandez, M.~Khramtsov and M.~Knysh,
  \emph{{Factorization of the Hilbert space of eternal black holes in general
  relativity}},  \href{https://arxiv.org/abs/2410.00091}{{\ttfamily
  2410.00091}}.

\bibitem{Kourkoulou17}
I.~Kourkoulou and J.~Maldacena, \emph{{Pure states in the SYK model and
  nearly-AdS2 gravity}},  \href{https://arxiv.org/abs/1707.02325}{{\ttfamily
  1707.02325}}.

\bibitem{Penington19}
G.~Penington, S.H.~Shenker, D.~Stanford and Z.~Yang, \emph{{Replica wormholes
  and the black hole interior}},
  \href{https://doi.org/10.1007/jhep03(2022)205}{\emph{Journal of High Energy
  Physics} {\bfseries 2022} (2022) }
  [\href{https://arxiv.org/abs/1911.11977}{{\ttfamily 1911.11977}}].

\bibitem{Chandra:2022fwi}
J.~Chandra and T.~Hartman, \emph{{Coarse graining pure states in AdS/CFT}},
  \href{https://doi.org/10.1007/JHEP10(2023)030}{\emph{JHEP} {\bfseries 10}
  (2023) 030} [\href{https://arxiv.org/abs/2206.03414}{{\ttfamily
  2206.03414}}].

\bibitem{Boruch:2023trc}
J.~Boruch, L.V.~Iliesiu and C.~Yan, \emph{{Constructing all BPS black hole
  microstates from the gravitational path integral}},
  \href{https://doi.org/10.1007/JHEP09(2024)058}{\emph{JHEP} {\bfseries 09}
  (2024) 058} [\href{https://arxiv.org/abs/2307.13051}{{\ttfamily
  2307.13051}}].

\bibitem{Boruch:2024kvv}
J.~Boruch, L.V.~Iliesiu, G.~Lin and C.~Yan, \emph{{How the Hilbert space of
  two-sided black holes factorises}},
  \href{https://arxiv.org/abs/2406.04396}{{\ttfamily 2406.04396}}.

\bibitem{Geng:2024jmm}
H.~Geng and Y.~Jiang, \emph{{Microscopic Origin of the Entropy of Single-sided
  Black Holes}},  \href{https://arxiv.org/abs/2409.12219}{{\ttfamily
  2409.12219}}.

\bibitem{Li2024}
P.~Li, \emph{{Notes on the Factorisation of the Hilbert Space for Two-Sided
  Black Holes in Higher Dimensions}},
  \href{https://arxiv.org/abs/2410.23886}{{\ttfamily 2410.23886}}.

\bibitem{Banerjee2024}
S.~Banerjee, J.~Erdmenger and J.~Karl, \emph{{Non-Locality induces Isometry and
  Factorisation in Holography}},
  \href{https://arxiv.org/abs/2411.09616}{{\ttfamily 2411.09616}}.

\bibitem{Bocchieri:1957cnr}
P.~Bocchieri and A.~Loinger, \emph{{Quantum Recurrence Theorem}},
  \href{https://doi.org/10.1103/PhysRev.107.337}{\emph{Phys. Rev.} {\bfseries
  107} (1957) 337}.

\bibitem{Percival:1961}
I.C.~Percival, \emph{Almost periodicity and the quantal h theorem},
  \href{https://doi.org/10.1063/1.1703705}{\emph{Journal of Mathematical
  Physics} {\bfseries 2} (1961) 235}
  [\href{https://arxiv.org/abs/https://pubs.aip.org/aip/jmp/article-pdf/2/2/235/19141887/235\_1\_online.pdf}{{\ttfamily
  https://pubs.aip.org/aip/jmp/article-pdf/2/2/235/19141887/235\_1\_online.pdf}}].

\bibitem{Schulman:1978}
L.S.~Schulman, \emph{Note on the quantum recurrence theorem},
  \href{https://doi.org/10.1103/PhysRevA.18.2379}{\emph{Phys. Rev. A}
  {\bfseries 18} (1978) 2379}.

\bibitem{Dyson:2002pf}
L.~Dyson, M.~Kleban and L.~Susskind, \emph{{Disturbing implications of a
  cosmological constant}},
  \href{https://doi.org/10.1088/1126-6708/2002/10/011}{\emph{JHEP} {\bfseries
  10} (2002) 011} [\href{https://arxiv.org/abs/hep-th/0208013}{{\ttfamily
  hep-th/0208013}}].

\bibitem{Sasieta:2022ksu}
M.~Sasieta, \emph{{Wormholes from heavy operator statistics in AdS/CFT}},
  \href{https://doi.org/10.1007/JHEP03(2023)158}{\emph{JHEP} {\bfseries 03}
  (2023) 158} [\href{https://arxiv.org/abs/2211.11794}{{\ttfamily
  2211.11794}}].

\bibitem{Goel:2018ubv}
A.~Goel, H.T.~Lam, G.J.~Turiaci and H.~Verlinde, \emph{{Expanding the Black
  Hole Interior: Partially Entangled Thermal States in SYK}},
  \href{https://doi.org/10.1007/JHEP02(2019)156}{\emph{JHEP} {\bfseries 02}
  (2019) 156} [\href{https://arxiv.org/abs/1807.03916}{{\ttfamily
  1807.03916}}].

\bibitem{Hamilton:2006az}
A.~Hamilton, D.N.~Kabat, G.~Lifschytz and D.A.~Lowe, \emph{{Holographic
  representation of local bulk operators}},
  \href{https://doi.org/10.1103/PhysRevD.74.066009}{\emph{Phys. Rev. D}
  {\bfseries 74} (2006) 066009}
  [\href{https://arxiv.org/abs/hep-th/0606141}{{\ttfamily hep-th/0606141}}].

\bibitem{Israel:1966rt}
W.~Israel, \emph{{Singular hypersurfaces and thin shells in general
  relativity}}, \href{https://doi.org/10.1007/BF02710419}{\emph{Nuovo Cim. B}
  {\bfseries 44S10} (1966) 1}.

\bibitem{Hawking:1982dh}
S.W.~Hawking and D.N.~Page, \emph{{Thermodynamics of Black Holes in anti-De
  Sitter Space}}, \href{https://doi.org/10.1007/BF01208266}{\emph{Commun. Math.
  Phys.} {\bfseries 87} (1983) 577}.

\bibitem{Henningson:1998gx}
M.~Henningson and K.~Skenderis, \emph{{The Holographic Weyl anomaly}},
  \href{https://doi.org/10.1088/1126-6708/1998/07/023}{\emph{JHEP} {\bfseries
  07} (1998) 023} [\href{https://arxiv.org/abs/hep-th/9806087}{{\ttfamily
  hep-th/9806087}}].

\bibitem{Balasubramanian:1999re}
V.~Balasubramanian and P.~Kraus, \emph{{A Stress tensor for Anti-de Sitter
  gravity}}, \href{https://doi.org/10.1007/s002200050764}{\emph{Commun. Math.
  Phys.} {\bfseries 208} (1999) 413}
  [\href{https://arxiv.org/abs/hep-th/9902121}{{\ttfamily hep-th/9902121}}].

\bibitem{Skenderis:2002wp}
K.~Skenderis, \emph{{Lecture notes on holographic renormalization}},
  \href{https://doi.org/10.1088/0264-9381/19/22/306}{\emph{Class. Quant. Grav.}
  {\bfseries 19} (2002) 5849}
  [\href{https://arxiv.org/abs/hep-th/0209067}{{\ttfamily hep-th/0209067}}].

\bibitem{Emparan:1999pm}
R.~Emparan, C.V.~Johnson and R.C.~Myers, \emph{{Surface terms as counterterms
  in the AdS / CFT correspondence}},
  \href{https://doi.org/10.1103/PhysRevD.60.104001}{\emph{Phys. Rev. D}
  {\bfseries 60} (1999) 104001}
  [\href{https://arxiv.org/abs/hep-th/9903238}{{\ttfamily hep-th/9903238}}].

\bibitem{Marolf:2018ldl}
D.~Marolf, \emph{{Microcanonical Path Integrals and the Holography of small
  Black Hole Interiors}},
  \href{https://doi.org/10.1007/JHEP09(2018)114}{\emph{JHEP} {\bfseries 09}
  (2018) 114} [\href{https://arxiv.org/abs/1808.00394}{{\ttfamily
  1808.00394}}].

\bibitem{Srednicki1994}
M.~Srednicki, \emph{{Chaos and quantum thermalization}},
  \href{https://doi.org/10.1103/PhysRevE.50.888}{\emph{Physical Review E}
  {\bfseries 50} (1994) 888} [\href{https://arxiv.org/abs/9403051}{{\ttfamily
  9403051}}].

\bibitem{Gibbons1977}
G.W.~Gibbons and S.W.~Hawking, \emph{{Action integrals and partition functions
  in quantum gravity}},
  \href{https://doi.org/10.1103/PhysRevD.15.2752}{\emph{Physical Review D}
  {\bfseries 15} (1977) 2752}.

\bibitem{Antonini:2024yif}
S.~Antonini, V.~Balasubramanian, N.~Bao, C.~Cao and W.~Chemissany,
  \emph{{Non-isometry, State-Dependence and Holography}},
  \href{https://arxiv.org/abs/2411.07296}{{\ttfamily 2411.07296}}.

\end{thebibliography}\endgroup

\end{document}